\documentclass[journal]{IEEEtran}

\usepackage{amsmath}
\usepackage{subfigure}
\usepackage{lineno, hyperref}
\usepackage{xcolor}
\usepackage{extarrows}
\usepackage{algorithm}
\usepackage{algorithmic}
\usepackage{setspace}
\usepackage{amssymb}
\usepackage{bm}
\usepackage{graphicx}
\usepackage[singlelinecheck=off]{caption}

\usepackage{graphicx}
\usepackage{times}
\usepackage{epsfig}
\usepackage{graphicx}
\usepackage{amsmath}
\usepackage{amssymb}
\usepackage{latexsym}
\usepackage{algorithmic}
\usepackage{array}
\usepackage{mdwmath}
\usepackage{mdwtab}
\usepackage{eqparbox}
\usepackage{bm}
\usepackage{mathtools}
\usepackage{multirow}
\usepackage{booktabs}
\usepackage{xcolor}
\usepackage{cases}
\usepackage{harpoon}
\usepackage{wrapfig}
\usepackage{cite}
\usepackage{flushend}
\usepackage{CJK}
\usepackage{indentfirst}
\usepackage{amsmath}
\usepackage{cases}
\usepackage{cuted}

\ifCLASSINFOpdf

\begin{document}
		%

		\title{RIS-Assisted Joint Uplink Communication and Imaging: Phase Optimization and Bayesian \\Echo Decoupling}  

		%
		%
		%
		
		\author{Shengyu~Zhu, Zehua~Yu, Qinghua~Guo, \IEEEmembership{Senior Member,~IEEE}, Jinshan~Ding, \IEEEmembership{Member,~IEEE}, \\      
			Qiang~Cheng, \IEEEmembership{Senior Member,~IEEE}, and Tie~Jun~Cui, ~\IEEEmembership{Fellow,~IEEE}
			\thanks{This work was supported by the Basic Scientific Center of Information Metamaterials of the National Natural Science Foundation of China under Grant 6228810001 and also supported by the National Natural Science Foundation of China under Grant 62171358.}
			\thanks{Shengyu Zhu and Jinshan Ding are with the National Laboratory of Radar Signal Processing, Xidian University, Xi'an 710071, China (e-mail: ding@xidian.edu.cn). }
			\thanks{Zehua Yu is with the School of Electronic Engineering, Xidian University, Xi'an 710071, China (e-mail: yuzehua@xidian.edu.cn). }			
			\thanks{Qinghua Guo is with the School of Electrical, Computer, and Telecommunications Engineering, University of Wollongong, Wollongong, Australia (e-mail: qguo@uow.edu.au). }
			\thanks{Qiang Cheng and Tie Jun Cui are with the State Key Laboratory of Millimeter Waves, Southeast University, Nanjing 210096, China (e-mail: tjcui@seu.edu.cn). }
	
		}

		\markboth{IEEE TRANSACTIONS ON SIGNAL PROCESSING , ~Vol. ~XX, No. ~XX, 2022}%
		{Shell \MakeLowercase{\textit{et al. }}: Bare Demo of IEEEtran. cls for IEEE Journals}
	

		\maketitle
		\begin{abstract} 
		 Achieving integrated sensing and communication (ISAC) via uplink transmission is challenging due to the unknown waveform and the coupling of communication and sensing echoes. In this paper, a joint uplink communication and imaging system is proposed for the first time, where a reconfigurable intelligent surface (RIS) is used to manipulate the electromagnetic signals for echo decoupling at the base station (BS). Aiming to enhance the transmission gain in desired directions and generate required radiation pattern in the region of interest (RoI), a phase optimization problem for RIS is formulated, which is high dimensional and nonconvex with discrete constraints. To tackle this problem, a back propagation based phase design scheme for both continuous and discrete phase models is developed.  Moreover, the echo decoupling problem is tackled using the Bayesian method with the factor graph technique, where the problem is represented by a graph model which consists of difficult local functions. Based on the graph model, a message-passing algorithm is derived, which can efficiently cooperate with the adaptive sparse Bayesian learning (SBL) to achieve joint communication and imaging. Numerical results show that the proposed method approaches the relevant lower bound asymptotically, and the communication performance can be enhanced with the utilization of imaging echoes.
		\end{abstract}
		

		%
		
		\IEEEpeerreviewmaketitle

\section{Introduction}
\IEEEPARstart{I}{ntegrated} sensing and communication (ISAC) has been recognized as a promising technology for the next-generation wireless networks\cite{9606831}. ISAC not only allows communication and radar systems to share spectrum resources, but also enables communication and radar sensing functionalities simultaneously, which significantly reduces hardware resource consumption. However, this new framework introduces great challenges to hardware design and signal processing, thus bringing a research upsurge in system design \citeonline{systemdesign1,systemdesign2,systemdesign3,systemdesign4,systemdesign5}, optimal joint waveform design \cite{caoJointRadarCommunicationWaveform2020,dongLowComplexityBeamformerDesign2021,liu28DualfunctionalRadarCommunication2018,luoOptimizationQuantizationMultibeam2019,shiConstrainedWaveformDesign2020}, exploration of joint processing algorithms \cite{hassanienSignalingStrategiesDualfunction2016,zhangOverviewSignalProcessing2021,zhengRadarCommunicationCoexistence2019}, and resource allocation\cite{chalise29PerformanceTradeoff2017,liu37SurveyFundamental2021,renPerformanceTradeoffsJoint2019}.

In the past years, information metamaterials have opened a new era of real-time digital regulation of electromagnetic waves \cite{cui2014coding, liu2017concepts,CUI2020101403,chenArtificialNeuralNetwork,daiSimultaneousSituDirection2022,huangSingleMetasurfaceCan2022, liHighResolutionNearFieldImaging2022b,wangIntelligentElectromagneticMetasurface2022,wanJointRadarCommunication2022,wanUserTrackingWireless2021a,weiwangHighPrecisionDirectionofArrivalEstimations2022,zhouTwodimensionalDirectionofarrivalEstimation2022}. As one representative of information metamaterials, reconfigurable intelligent surface (RIS) is expected to break the dependence of traditional sensing and communication on the channel environment. RIS bridges the gap between ISAC and information metamaterials by configuring the electromagnetic environment more flexibly, quickly and intelligently, which is highly expected to better serve the next generation of wireless communication and senisng. Essentially, it is a low-cost passive system that achieves control of electromagnetic beam energy by tuning the phase of array elements. It is shown in \cite{cui2014coding} that the highly controllable reflection of RIS can be practically achieved by leveraging the existing digitally reconfigurable or programmable metasurface. RIS was first applied as a wireless relay to redirect electromagnetic signals in the field of wireless communication \cite{zhongji1,zhongji2,zhongji3}, thereby overcoming the adverse effects in natural environments. In addition, it can also be used as a new communication architecture transmitter to implement various modulation \cite{zhao14FSKProgrammable2019,direct1,direct2,dai16QAKRealization2020,liNovelWirelessCommunication,tang19MIMOTransmission2020,wan18ASKMultichannel2019,zhang17WirelessCommunication2021}, e.g.,  FSK \cite{zhao14FSKProgrammable2019}, PSK \cite{direct2}, and MIMO \cite{tang19MIMOTransmission2020}. Recently, RIS has been employed to achieve environment awareness and parameters estimation. 
A new self-sensing RIS architecture was proposed in \cite{42}, where the performance of different benchmark sensing systems in the cases of with and without RIS was compared.
The authors in \cite{54} and \cite{55} shed light on the interplay among the system parameters, including the radar-RIS distance, the RIS size and the location of the prospective target. The results show that the radar system can achieve the optimal performance when the RIS is deployed in the near field of the radar arrays on both the transmitter and receiver sides. 
The authors in \cite{10} and \cite{dai2020reconfigurable} proposed to sense humans, recognize their gestures and physiological state simultaneously by utilizing the programmable metasurface and Wi-Fi signals. 

Recently, RIS-assisted ISAC system has attracted significant research interest. A Dual-Functional Radar-Communication (DFRC) system with RIS deployed near the communication devices has been proposed in \cite{40,41}. By optimizing the Cramer-Rao bound (CRB) of DOA estimation, the constant-waveform and RIS phase shifts are designed jointly to mitigate the mitigating multi-user interference (MUI). The RIS-assisted ISAC system in the cases of congested and obstructed channels has been investigated in \cite{44} and \cite{48}, respectively, where the PSM of the RIS and the precoding of the base station (BS) are optimized jointly to improve the SNR at receiver. 
To explore the potential of multiple RISs in assisting ISAC, the authors in \cite{43} propose a double-RIS-assisted ISAC system, where two RISs are deployed to enhance the communication performance while suppressing mutual interference. The beam patterns of RISs and radar are optimized jointly based on penalized dual decomposition (PDD) to enhane the system performance. In \cite{Location_Awareness}, the authors propose an RIS-aided localization and communication system, where the theoretical performance limits of localization and communication is derived for both near-field and far-field scenarios. Numerical results show that with the assistance of multiple RISs, both spectral efficiency and localization accuracy can be improved significantly.

Most of the existing literature in ISAC focus on the utilization of downlink transmission. On the contrary, ISAC in uplink transmission is underexplored, technically because it is much more challenging compared with downlink counterparts due to unknown signal waveform and coupling of communication and sensing echoes. In this article, we propose a new RIS-assisted ISAC system, where the uplink transmission is exploited to achieve joint communication and imaging. In contrast to existing works \cite{40,41,44,48,43} that achieve communication and sensing at different receivers, the proposed uplink ISAC system performs communication and imaging at one receiver, where the communication performance can be enhanced with the utilization of sensing echo. We show that the radiation pattern can be controlled expectedly with the proposed phase design scheme and the factor graph technique is very suitable for echo decoupling for uplink ISAC system. 
The contribution of this work can be summarized as follows

\begin{itemize}
 \item A novel RIS-assisted uplink joint communication and imaging system is presented for the first time, to the best of our knowledge. It is shown that by properly designing the phase shift of the RIS, the radiation pattern can be modulated as desired, thereby allowing joint communication and imaging at the BS with only one RF chain. 

 \item A phase optimization problem based on the requirements of the system is formulated. A back propagation based phase design scheme with the combination of temperature parameter is proposed to achieve efficient phase optimization for both continuous and discrete phase models. 

 
  
 \item A factor graph representation of the joint communication and imaging is established, and then, an efficient message passing algorithm is successfully developed to decoupling echoes at the BS. It is also demonstrated that the communication performance can be enhanced by making full use of imaging echoes. 
 
\end{itemize}			

The article is organized as follows. Section II briefly introduces the system and the signal model used. In Section III, the phase design problem is formulated and a back propagation based optimization method is proposed. In Section IV, a joint maximum a posteriori estimation problem is established. Then the problem is represented by a graph model, based on which we propose an efficient message passing algorithm to achieve joint communication and imaging. Numerical results are provided in Section V. Section VI concludes this article. 
		
Notations: Throughout this paper, column vectors and matrices are denoted by bold lower-case and bold upper-case letters, respectively. The notation $(\cdot)^{\mathrm H}$ denotes the conjugate operator and $(\cdot)^{-1}$ denotes inverse operator. $|\cdot|$, $\|\cdot\|$ and $\|\cdot\|_{\mathrm F}$ are the $l_2$ norm, the $l_2$ norm and the Frobenius norm, respectively. The transpose operation is denoted by $(\cdot)^{\mathrm T}$.  We use $\mathbb{E}\{\cdot\}$ to denote the expectation operator. The notation $\mathcal{N}(\textbf{x};\mathbf{m}, \mathbf{V})$ denotes a Gaussian distribution of $\textbf{x}$ with mean vector $\mathbf{m}$ and covariance matrix $\mathbf{V}$. In many cases, we also use the inverse of the covariance matrix $\mathbf{V}$, which is denoted by $\mathbf{W}$ and called weight matrix as in \cite{loeliger2004an,procdi_facgph}.  The notation $\mathbf{I}_{a}$ denotes an identity matrix of a size $a \times a$. The notation $\propto$ denotes equality of functions up to a scale factor. We use 
$\odot$ to represent the inner product, and $\int_{\sim a} \mathcal{F} (\cdot)$ denotes integral over all variables in $\mathcal{F} (\cdot)$ except $a$. The notation $\mathrm{diag(\cdot)}$ denotes the diagonal operation and $\delta(\cdot)$ is the Dirac delta function.

		\section{System Model}
		
		\begin{figure}
			\centering
			\includegraphics[width=0.9\linewidth]{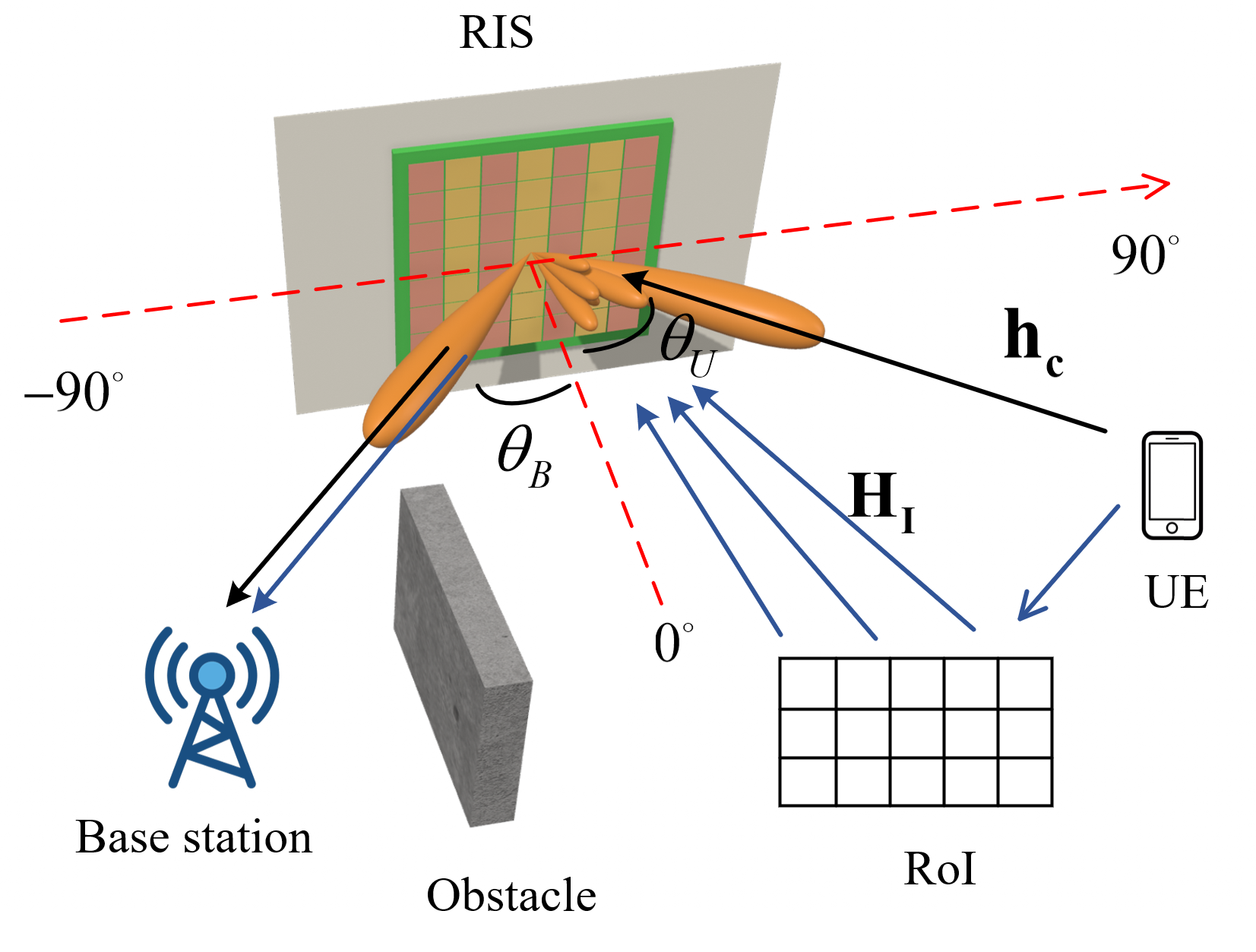}
			\caption{Diagram of a RIS-assisted joint uplink imaging and communication system.}
			\label{fig1}
		\end{figure}

We consider an uplink RIS-assisted ISAC system in 2D plane as shown in Fig.\ref{fig1}, which consists of a user equipment (UE), a BS, a RIS and a region of interest (RoI). Assume there is an obstacle between the UE and the BS. The RIS is deployed between the UE and BS to customize the radio environment for the signal from the UE side and reflect to the BS, and its center is considered as the origin of the coordinate system. The UE sends communication signals to the BS with the assistant of the RIS. Due to the high path loss, it is also assumed that the signals with more than two reflections are negligible. We aim to perform communication at the BS while realizing reconstruction of the RoI.
     
The UE and the BS are located at the far field of the RIS, in directions of ${\theta _U}$ and ${\theta _B}$ from the origin, respectively, and both of which are assumed to adopt the single-antenna structure. The RIS is a uniform linear array (ULA) with $N$ elements that are half a wavelength apart. The RoI is divided into an equi-spaced grid with $M$ pixel units. We denote the angle vector of the RoI from the origin as $\boldsymbol{\theta}_S=\left[\theta_{S, 1}, \theta_{S, 2}, \ldots, \theta_{S, M} \right]^T$. 

\begin{figure}
        \centering
    	\includegraphics[width=0.7\linewidth]{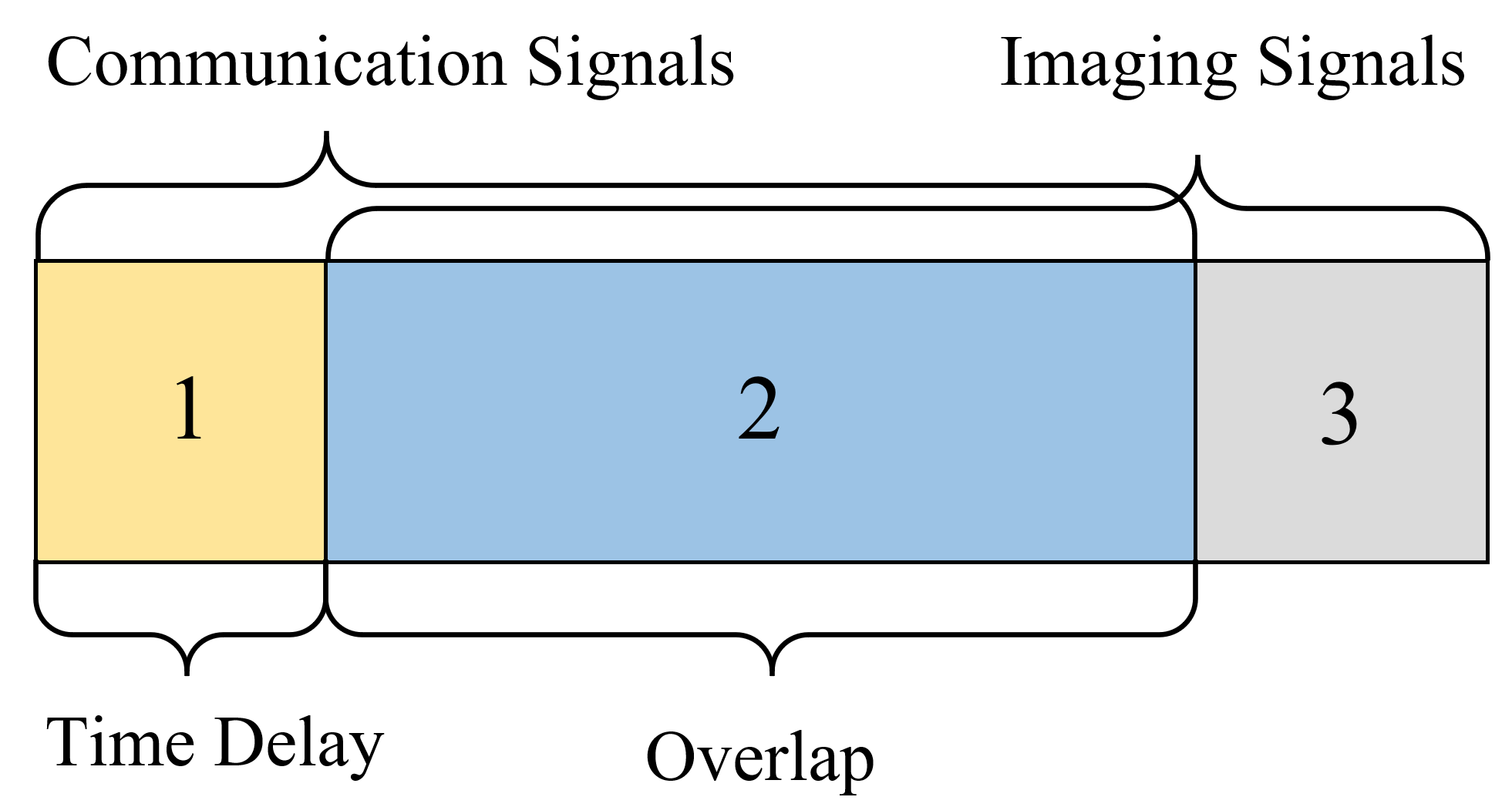}
    	\caption{Diagram of overlapped receive signal model. }
    	\label{fig:2}
    \end{figure}

It is noted that at the transmitted signals reach the BS through two paths, i.e., UE-RIS-BS, namely communication link and UE-RoI-RIS-BS, namely imagine link. Supposing that the UE transmits a communication frame of length $L$, due to the difference of propagation delay, the signal received at the BS falls into 3 categories: 1) non-overlapped communication echo, 2) overlapped communication and imaging echo and 3) non-overlapped imaging echo. 
A generic model of received signal is shown in Fig.\ref{fig:2}. Specifically, the sampled base band received signal at time index $t$ can be expressed as 		
		\begin{equation} \label{recv_sig}
			\begin{aligned}
				y(t)\! =\! \left\{ \! \! {\begin{array}{l}
						{\underbrace {{\alpha_c}{{\bf{g}}^{\rm{T}}}{\bf{\Theta }}(t){{\bf{h}}_{\bf{c}}}x(t)}_{UE \to RIS \to BS} + w(t), \quad t \leq k} \\ \\
						{\underbrace {{\alpha_c}{{\bf{g}}^{\rm{T}}}{\bf{\Theta }}(t){{\bf{h}}_{\bf{c}}}x(t)}_{UE \to RIS \to BS} + \underbrace {{\alpha_I}{{\bf{g}}^{\rm{T}}}{\bf{\Theta }}(t){{\bf{H}}_{\bf{I}}}{\boldsymbol{\sigma }}x(t - k)}_{UE \to RoI \to RIS \to BS} + w(t)}, \\ \qquad \qquad \qquad \qquad \qquad \qquad \qquad \qquad {k < t < L}\\
						{\underbrace {{\alpha_I}{{\bf{g}}^{\rm{T}}}{\bf{\Theta }}(t){{\bf{H}}_{\bf{I}}}{\boldsymbol{\sigma }}x(t - k)}_{UE \to RoI \to RIS \to BS} + w(t), \quad L < t \leq L + k}
				\end{array}} \right.
			\end{aligned}
		\end{equation}
		where we define the variables and symbols list below
		\begin{itemize}
			\item $\alpha_c$ and $\alpha_I$ denote the attenuation coefficients of communication and imaging links, respectively, which are assumed to have been obtained by the pilot signals.			
			\item The steering vectors are ${\bf{g}} = \mathbf{a}({\theta _B}) \in {\mathbb{C}^{N \times 1}}$, ${{\bf{h}}_{\bf{c}}} = \mathbf{a} ({\theta _U}) \in {\mathbb{C}^{N \times 1}}$ and ${{\mathbf{H}}_{\bf{I}}} = \mathbf{a} (\boldsymbol{\theta}_S) \in {\mathbb{C}^{N \times M}}$, where ${\mathbf{a}}(\theta ) = {\left[ {1, {e^{j{{\pi }}\sin (\theta )}}, \ldots , {e^{j{{\pi }}\left( {N - 1} \right)\sin (\theta )}}} \right]^T}$.
			\item ${{\bf{\Theta }}(t) = {\rm{diag}}\left( {{e^{ j {\theta _{t,1}}}}, \ldots ,{e^{j {\theta _{t,N}}}}} \right)}$ denotes the phase shift matrix of RIS at the time index $t$, where 		
			${\theta _{t,n}} \in \{ 0,\frac{{2\pi }}{{2^{n_{bit}}}},...,\frac{{2\pi(2^{n_{bit}}-1) }}{{2^{n_{bit}}}} \}$, $n_{bit}$ denotes the number of discrete values \cite{di2020reconfigurable, dai2020reconfigurable}. When $n_{bit}\to +\infty$, it means that the phase shift of RIS can be controlled continuous. 			
			\item $k$ indicates the time delay between communication and imaging links, which is assumed to has been estimated by the pilot signals \cite{pilotsignal}.
			\item $x(t)$ denotes the communication symbol sent by the UE. In this paper, we consider QPSK codes for communication, i.e.,  $x(t) \in \{\exp(j\frac{\pi}{4}),\exp(j\frac{3\pi}{4}),\exp(j\frac{5\pi}{4}), \exp(j\frac{7\pi}{4}) \}$.			
			\item $\boldsymbol{\sigma} \in {\mathbb{C}^{M \times 1}} $ denotes the vector of scattering coefficients of the RoI, which is assumed to be sparse.
			\item $w(t)$ is the i.i.d. complex Gaussian noise with zero mean and variance $\xi^2$.		
		\end{itemize} 
      
\section{Phase Shift Design Scheme}
In this section, a phase optimization problem is established based on the requirement of the joint communication and imaging system. We then present a back propagation based phase design scheme in the both cases of continuous and discrete phase models. 

\subsection{Phase Shifts Optimization Formulation}
The phase shift $\left\{\mathbf{\Theta}(t) \right\} $ is critical since it determines the gain of UE-RIS-BE and RoI-RIS-BS links simultaneously, and both communication and imaging performance should be considered in optimizing $\left\{\mathbf{\Theta}(t) \right\} $. For communication functionality, we hope to reflect more radiated energy from the UE to the BS after modulation by the RIS, i.e., maximizing ${\bf{g}}^{\rm{T}}{\bf{\Theta }}(t){{\bf{h}}_{\bf{c}}}$. For imaging functionality, on the one hand, we hope to gather more radiated energy reflected by the RoI and reflect it to the BS, i.e., maximizing ${{{\bf{g}}^T}{\bf{\Theta }}(t){{\bf{H}}_{\bf{I}}}}$. On the other hand, according to the compressed sensing theory \cite{CS1,CS2}, when reconstructing $\boldsymbol{\sigma}$, we need to modify $\left\{\mathbf{\Theta}(t), k<t \leq L+k \right\} $ to minimize the correlation between columns of the sensing matrix which given by 
\begin{equation}\label{G}
	\begin{array}{*{20}{l}}
		{\bf{G}} = \left[ {\begin{array}{*{20}{c}}
				{{{\bf{g}}^T}{\bf{\Theta }}(k){{\bf{H}}_{\bf{I}}}}\\
				{\vdots}\\
				{{{\bf{g}}^T}{\bf{\Theta }}(L+k){{\bf{H}}_{\bf{I}}}}
		\end{array}} \right].
	\end{array}
\end{equation}

The orthogonality of $\bf{G}$ can be evaluated by the mean of non-diagonal elements of $\mathcal{R} (\bf{G})$, i.e., $\left\| {{\mathcal{ R}} ({\bf{G}}) - {\bf{I}}_M} \right\|_{\mathrm F}/(M^2-M)$, where $\mathcal{R} (\bf{G})$ denotes the correlation coefficient matrix of columns of $\bf{G}$. Based on the above consideration, we can formulate the following phase shift optimization problem
\begin{subequations} \label{const}
\begin{align} 
		& \! \! \!\! \! \! \! \! \! \! \! \max_{\boldsymbol{\Theta }(t)} 
		\sum\limits_{t = 1}^{L + {\rm{k}}} {\rho {{ {\left\| {{{\bf{g}}^T}{\bf{\Theta }}(t){{\bf{h}}_{\bf{c}}}} \right\|} }^2} + (1 - \rho ){{ {\left\| {{{\bf{g}}^T}{\bf{\Theta }}(t){{\bf{H}}_{\bf{I}}}} \right\|} }^2}} \label{const_1}  \\ \text { s.t. } 
		&\frac{\left\| {{\mathcal{ R}} ({\bf{G}}) - {\bf{I}}_M} \right\|_{\mathrm F}}{M^2-M} \le {\eta_0},   \label{const_1a} \\ 
		&{{\theta _{t,n}} \in \left\{ 0,\frac{{2\pi }}{{{2^{{n_{bit}}}}}},...,\frac{{2\pi ({2^{{n_{bit}}}} - 1)}}{{{2^{{n_{bit}}}}}} \right\} }.  \label{const_1b}
\end{align}
\end{subequations}
The cost function \eqref{const_1} indicates that we expect to concentrate energy in the RoI and the direction of communication as much as possible, where $\rho \in[0,1]$ is used to realize a trade-off between communication and imaging performance. ${\eta_0}$ is a threshold between 0 and 1 to constrain the orthogonality of $\bf{G}$. The constraint \eqref{const_1b} denotes the feasible set of phase shift. 
We noted that \eqref{const} is non-convex, highly nonlinear and there are numerous coupled variables to be optimized, making it highly challenging to solve the problem directly.

\subsection{Back Propagation Based Phase Shift Optimization }
\begin{figure}
	\centering
	\includegraphics[width=0.95\linewidth]{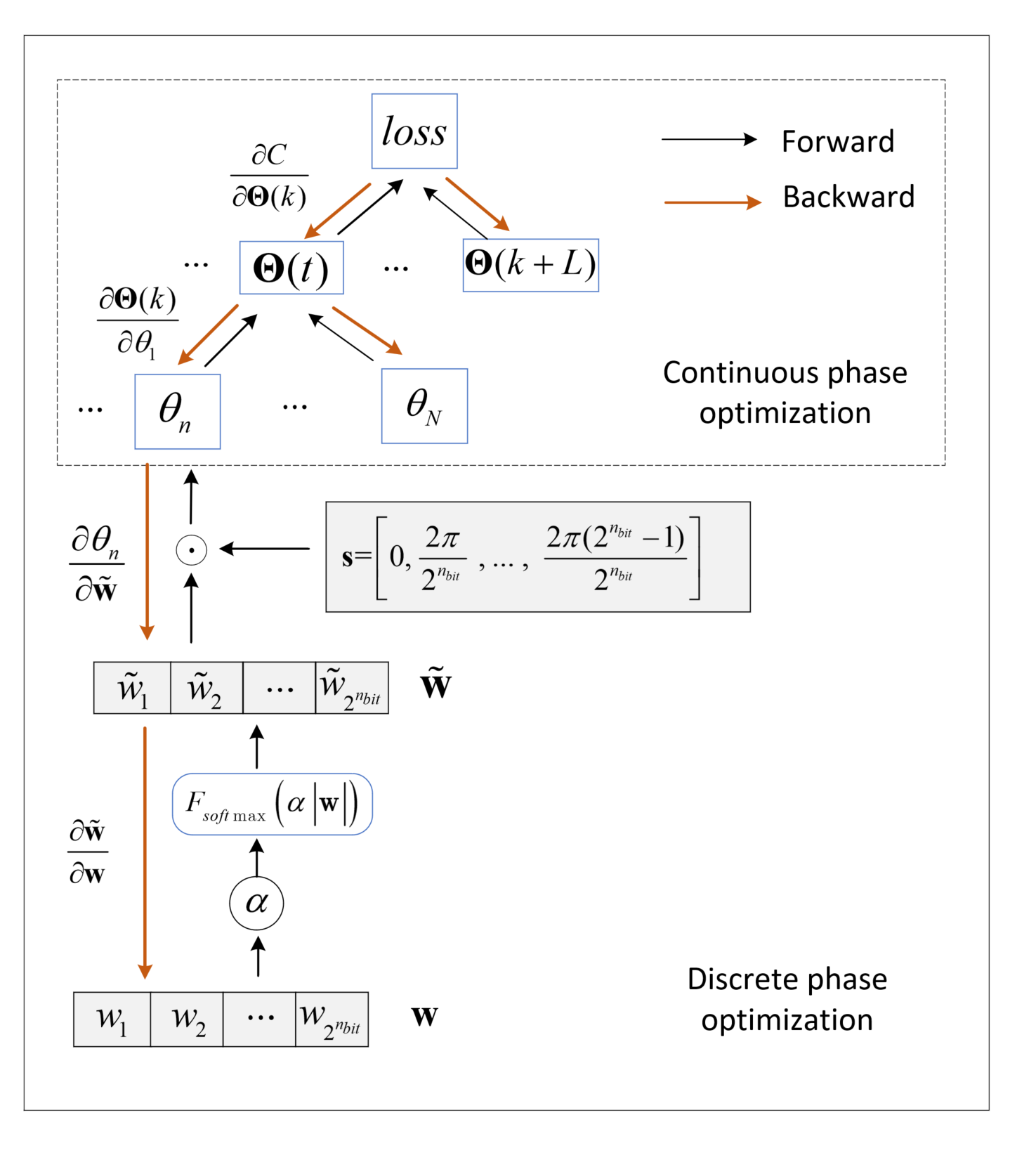}
	\caption{The flowchart of the proposed phase optimization method for continuous and discrete phase model.}
	\label{BP_fig}
\end{figure}

Note that \eqref{const_1a} indicates a strongly coupled constraint on $\left\{\mathbf{\Theta}(t), 1<t \leq L+k \right\} $, while the constraint \eqref{const_1b} makes the problem even more challenging. To solve this problem, we relax the original problem \eqref{const} into two subproblems and turn to find the sub-optimal solution. Specifically, we propose to first initialize $\left\{\mathbf{\Theta}(t), 1\leq t \leq L+k \right\} $ by minimizing $\left\| {{\mathcal{ R}} ({\bf{G}}) - {\bf{I}}_M} \right\|/(M^2-M)$ subject to the constraint \eqref{const_1b}. With an appropriate start point of $\left\{\mathbf{\Theta}(t), 1\leq t \leq L+k \right\}$, we can further modify it to maximize \eqref{const_1} until \eqref{const_1a} and \eqref{const_1b} can not be satisfied. With this in mind, we first need to solve the following problem
\begin{align}  \label{sub_prob1}
   		\min_{\boldsymbol{\Theta }(t)}
   		&\ \frac{\left\| {{\mathcal{ R}} ({\bf{G}}) - {\bf{I}}_M} \right\|_{\mathrm F}}{M^2-M}
      \\  \text { s.t. } & \eqref{const_1b}.	\notag
\end{align}
However, the above problem is still challenging due to the coupled cost function.  Since Gaussian matrix is recognized to be suit for a sensing matrix \cite{Gaussin_sensing}, we convert \eqref{sub_prob1} into the following problem 
\begin{align} \label{opt221}
   	 \min_{\boldsymbol{\Theta }(t)}&  \ \  {\left\| {\mathbf{R}_{L\times N} - {\bf{ G}}}\right\|_{\rm{F}}}  \\  \text { s.t. } & \eqref{const_1b}.   \notag
\end{align}
where $\mathbf{R}_{L\times N}$ is a Gaussian matrix in the size of $L\times N$. 

Inspired by the color section method in \cite{color_section}, we introduce an increasing temperature parameter $\alpha$ and the softmax function into back propagation framework to solve \eqref{opt221}, where the temperature parameter in the $l$th iteration is calculated by $\alpha  = 1 + {(rl)^2}$ and $r$ is a factor to adjust the increasing rate, 
and the softmax operation of $\alpha | \mathbf{w}_n^{l} |$ is defined as 
\begin{equation}
F_{soft\max} (\alpha | \mathbf{w}_n^{l} |) = \frac{{\exp (\alpha | \mathbf{w}_n^{l}|)}}{\sum_{s=1}^{{2^{{n_{bit}}}}} {\exp (\alpha \left| {w_{n,s}^{l}} \right|)}}
\end{equation}

 The corresponding flowchart of the proposed method is shown in Fig. \ref{BP_fig}, where $\left\{\mathbf{\Theta}(t) \right\}$ can be updated by alternating iterative forward and backward propagation procedures. In the forward propagation, assume that $\mathbf{w}_n$ in the $(l-1)$th iteration has been obtained, which denoted by $\mathbf{w}_n^{l-1}$, then $\mathbf{\tilde w}_n^l$ is calculated by making softmax operation for $\alpha | \mathbf{w}_n^{l-1} |$. 
 It is noteworthy that $\alpha$ increases with the iteration, and the softmax operation essentially plays a role of selector. Specifically, the softmax operation will push the larger element in $\mathbf{w_n}$ approaching 1 while the smaller elements approaching 0 with the increase of $\alpha$. 
 Then $ \theta_n $ can be obtained by making inner product of $\mathbf{\tilde w}_n^l$ and $\mathbf{s}$. After applying the above procedure for all $\{ \theta \} $, we can update $\left\{\mathbf{\Theta}(t) \right\} $ in parallel thereby calculate the loss in the $l$th iteration. In the backward propagation, the chain rule is utilized to update the gradients of all variables except $\mathbf{s}$ and thereby update $\mathbf{w}$ by stochastic gradient descent (SGD) \cite{bottou2018optimization}. We can finally obtain $\left\{\mathbf{\Theta}(t)\right\} $ along with corresponding $\mathbf{w}$ after the forward/backward propagation iterated alternately until convergence. The steps of the proposed method for solving \eqref{opt221} is summarized in Algorithm 1.  

Based the initialized $\left\{\mathbf{\Theta}(t) \right\}$ , we need to further modify $\left\{\mathbf{\Theta}(t) \right\} $ by solving the following problem
\begin{align} 
   		\min_{\boldsymbol{\Theta }(t)}
   		&\sum\limits_{t = 1}^{L + {\rm{k}}} \frac{1}{{\rho {{ {\left\| {{{\bf{g}}^T}{\bf{\Theta }}(t){{\bf{h}}_{\bf{c}}}} \right\|} }^2} + (1 - \rho ){{ {\left\| {{{\bf{g}}^T}{\bf{\Theta }}(t){{\bf{H}}_{\bf{I}}}} \right\|} }^2}}} \label{opt222}
      \\  \text { s.t. } & \eqref{const_1b}.	\notag
\end{align}
Note that \eqref{const_1} is rewritten to be a minimized form to facilitate the calculation of loss.
Fortunately, it is also worth noting that \eqref{opt222} can be solved in the same way as \eqref{sub_prob1}. All we need to do is execute Algorithm 1, where the loss is calculated by the cost function in \eqref{opt222} and the iteration condition $\Gamma$ is given by \eqref{const_1a}. 

In practical, when the phase shift can be controlled continuously, i.e., $n_{bit}\to +\infty$, the constraint \eqref{const_1b} can be removed from \eqref{const}, which makes \eqref{const} greatly simplified. In this case, we can first initialize $\mathbf{\Theta}$ by minimizing $\left\| {{\mathcal{ R}} ({\bf{G}}) - {\bf{I}}_M} \right\|/(M^2-M)$. Then $\left\{\mathbf{\Theta}(t) \right\}$ can be further optimized by minimize $\sum\limits_{t = 1}^{L + {\rm{k}}} \frac{1}{{\rho {{ {\left\| {{{\bf{g}}^T}{\bf{\Theta }}(t){{\bf{h}}_{\bf{c}}}} \right\|} }^2} + (1 - \rho ){{ {\left\| {{{\bf{g}}^T}{\bf{\Theta }}(t){{\bf{H}}_{\bf{I}}}} \right\|} }^2}}}$,
which can be solved efficient by back propagation procedure in the dashed box of Fig.\ref{BP_fig}.

\begin{algorithm}[t]
	\caption{Discrete Phase Optimization Algorithm}
	\begin{algorithmic}[1]
		\STATE  Initialize the parameters: weights $\mathbf{w}_{t,n}^{1}$, learning rate $\eta$ and scale factor $r$
		\WHILE {$\Gamma$ }
		\STATE  $\alpha  = 1 + {(rl)^2}$
		\STATE  ${\forall n},{{\bf{\tilde w}}_{t,n}^{l}} = soft\max (\alpha | {{{\bf{w}}_{t,n}^{l}}} |)$
		\STATE  ${\forall n},{{\theta_{t,n}^{l}}} = {{\bf{\tilde w}}_{t,n}^{l}} \odot [0,\frac{{2\pi }}{{{2^{{n_{bit}}}}}},...,\frac{{2\pi ({2^{{n_{bit}}}} - 1)}}{{{2^{{n_{bit}}}}}}]$
		\STATE ${{\bf{\Theta }}^l(t) = {\rm{diag}}\left( {{e^{j {\theta_{t,1}^{l}}}}, \ldots ,{e^{j {\theta_{t,N}^{l}}}}} \right)}$
		\STATE  $C_t^l = Cost \ function({\bf{\Theta }}^l(t))$
		\STATE $\mathbf{W}_t^{l+1} = SGD \left( \mathbf{W}_t^{l}, \eta , \frac{\partial C_t^l} {\partial \mathbf{W}_t^l } \right)$ 
		\STATE $l = l + 1$
		\ENDWHILE
	\end{algorithmic}
	\label{Alg_2}
\end{algorithm}

\section{Message Passing Based Joint Communication and Imaging}

 In this part, we investigate joint communication and imaging at the BS. The problem is first formulated into a maximum posteriori probability problem, which is nonconvex and strongly coupled. To solve it, a factor graph representation is then established, based on which an efficient iterative message passing algorithm is derived to estimate $\boldsymbol{\sigma}$ and $\mathbf{x}$ jointly .

\subsection{Joint Communication and Imaging Formulation}

According to the Bayes' theorem, the joint posterior distribution of $\mathbf{x}$ and $\boldsymbol{\sigma}$ conditioned on $\mathbf{y}$ is given by
\begin{equation}\label{jt_distb}
  \begin{aligned}
	p(\mathbf{x},\boldsymbol{\sigma}|\mathbf{y}) &\propto p(\mathbf{y}|\mathbf{x},\boldsymbol{\sigma})p(\boldsymbol{\sigma})  p(\mathbf{x}) \\	
 \end{aligned}
\end{equation}
where $\mathbf{y} = [y(1),...,y(L+k)]^{\mathrm T} \in \mathbb{C}^{(L+k) \times 1} $ and $\mathbf{x} = [x(1),...,x(L)]^{\mathrm T} \in \mathbb{C}^{L \times 1}$; $p(\mathbf{x})$ and $p(\boldsymbol{\sigma})$ denote the prior for $\mathbf{x}$ and $\boldsymbol{\sigma}$, respectively. For QPSK modulation, we have
\begin{equation}\label{piror}
  	 p(\mathbf{x}) = \prod_{t=1}^{L} p(x(t)) = \prod_{t=1}^{L}\frac{1}{4}\sum_{i=1}^{4} \delta [ x(t)- e^{j(\frac{\pi}{2}i - \frac{\pi}{4})}].
\end{equation}                                                                                  
Following \cite{luo2021unitary}, it is assumed that the elements in $\boldsymbol{\sigma}$ are independent and following the two-layer sparsity-promoting prior, i.e.,
\begin{equation} \label{two_layer_prior}
	\begin{aligned}		
	p(\boldsymbol\sigma)  &= p(\boldsymbol\sigma|\bm{\gamma}) p(\boldsymbol\gamma|\epsilon, \eta)\\
	& = \prod_{m=1}^{M} \mathcal N \left( \sigma_m | 0, \gamma_m^{-1} \right) \cdot \prod_{m=1}^{M} Ga(\gamma_m^{-1}|\epsilon, \eta),
	\end{aligned}
\end{equation}                                                                  
where the precision vector $\bm{\gamma}=[\gamma_1,...,\gamma_M]^{\mathrm H}$. 

Based on \eqref{recv_sig}, $p(\mathbf{y}|\mathbf{x},\boldsymbol{\sigma})$ can be decomposed into the product of three categories of probability, i.e.,
\begin{equation}
  \begin{aligned}
	p(\mathbf{y}|\mathbf{x},\boldsymbol{\sigma})&\! =\! \prod_{t=1}^{ k} p\left(y(t)|x(t)\right) 
    \! \cdot\! \prod_{t=k+1}^{L}  p\left((y(t)|x(t),x(t-k),\boldsymbol{\sigma}\right) \\
	&\cdot \! \prod_{t=L+1}^{L+k}  p\left(y(t)|x(t-k),\boldsymbol{\sigma}\right) ,
 \end{aligned}                  
\end{equation}		
where
\begin{equation} \label{pyt1}
\begin{aligned}
  \begin{small}
 p\left(y(t)|x(t)\right) \propto  \exp \left\{ - \frac{\left[ y(t)- {\alpha_c}{{\bf{g}}^{\rm{T}}}{\bf{\Theta }}(t){{\bf{h}}_{\bf{c}}}x(t) \right]^2}{\xi^2}   \right\},  t\leq k 
  \end{small}
\end{aligned}
\end{equation}	    
\begin{equation}\label{pyt2}
	\begin{small}
	\begin{aligned}
	& p(y(t)|x(t),x(t-k),\boldsymbol{\sigma})\propto \\ & \exp \left\{ - \frac{\left[ y(t)- {\alpha_c}{{\bf{g}}^{\rm{T}}}{\bf{\Theta }}(t){{\bf{h}}_{\bf{c}}}x(t) - {\alpha_I}{{\bf{g}}^{\rm{T}}}{\bf{\Theta }}(t){{\bf{H}}_{\bf{I}}}{\boldsymbol{\sigma }}x(t - k) \right]^2}{\xi^2}   \right\},\\  & \qquad \qquad \qquad \qquad \qquad \qquad \qquad \qquad \qquad \qquad \qquad k< t\leq L  
    \end{aligned}
\end{small}
\end{equation}	 
\begin{equation}\label{pyt3}
	\begin{small}
	\begin{aligned}
		 p(&y(t)|x(t-k),\boldsymbol{\sigma}) \propto  \exp \left\{ - \frac{\left[ y(t) - {\alpha_I}{{\bf{g}}^{\rm{T}}}{\bf{\Theta }}(t){{\bf{H}}_{\bf{I}}}{\boldsymbol{\sigma }}x(t - k) \right]^2}{\xi^2}   \right\}, \\ & \qquad \qquad \qquad \qquad  \qquad \qquad \qquad \qquad \qquad \qquad  L<t\leq L+k  
	\end{aligned}
   \end{small}
\end{equation}	

Based on the decomposition in \eqref{piror} -  \eqref{pyt3}, we aim to find $\mathbf{x}$ and $\boldsymbol{\sigma}$ that maximize the joint posterior distribution $p(\mathbf{x},\boldsymbol{\sigma}|\mathbf{y})$ in \eqref{jt_distb}. However, it is a challenging problem as $\boldsymbol{\sigma}$ and $\mathbf{x}$ is strongly coupled in different time segments.

  \subsection{Factor Graph Representation for Joint Imaging and Communication}			
	
	\begin{table}
		\renewcommand\arraystretch{1.2}
		\centering
		\caption{Definitions of Variables and Functions}		
		\begin{tabular}{l} 
			\midrule
			$a_t^{} \buildrel \Delta \over = {\alpha _c}{{\bf{g}}^{\rm{T}}}{\bf{\Theta }}(t){{\bf{h}}_{\bf{c}}}x_t$ \\
			\hline
			${b_t} \buildrel \Delta \over = {{\alpha _I}{{\bf{g}}^{\rm{T}}}{\bf{\Theta }}(t)}{{\bf{H}}_{\bf{I}}}x_{t - k}$ \\
			\hline
			${c_t} \buildrel \Delta \over = a_t + {b_t}$ \\
			\hline
			${h_t} \buildrel \Delta \over = {\alpha _I}{{\bf{g}}^{\rm{T}}}{\bf{\Theta }}(t + k){{\bf{H}}_{\bf{I}}}\boldsymbol{\sigma}$ \\
			\hline
			$f_{A,t}^1 (a_t) \buildrel \Delta \over = {\cal N}(y_t; a_t^{},{\xi^2})$  \\
			\hline
			$f_{A,t}^2 (c_t) \buildrel \Delta \over = {\cal N}(y_t; c_t^{},{\xi^2})$  \\
			\hline
			$f_{A,t}^3 (b_t) \buildrel \Delta \over = {\cal N}(y_t; b_t^{},{\xi^2})$  \\
			\hline
			$f_{B,t} \left(a_t,x_t \right) \! \buildrel \! \Delta \over = \delta ({\alpha _c}{{\bf{g}}^{\rm{T}}}{\bf{\Theta }}(t){{\bf{h}}_{\bf{c}}}x_t - a_t^{})$  \\
			\hline
			${f_{D,t}}(a_t,b_t,c_t) \buildrel \Delta \over = \delta (a_t^{} + {b_t} - {c_t})$  \\
			\hline
			${f_{E,t}}\left(x_t \right) \buildrel \Delta \over = \frac{1}{4}\sum_{i=1}^{4} \delta [ x_t- e^{j(\frac{\pi}{2}i - \frac{\pi}{4})}]$   \\
			\hline
			${f_{K,t}}\left(h_t,x_t,b_{t+k} \right) \buildrel \Delta \over = \delta ({h_t}x_t - {b_{t+k}})$ \\
			\hline
			${f_{H,t}}(\boldsymbol{\sigma},h_t) \buildrel \Delta \over = \delta ({\alpha _I}{{\bf{g}}^{\rm{T}}}{\bf{\Theta }}(t+k){{\bf{H}}_{\bf{I}}}{\boldsymbol{\sigma}} - {h_t})$ \\
			\hline
			${f_G} (\boldsymbol{\sigma},\boldsymbol\gamma)\buildrel \Delta \over = \prod_M^{} {{\cal N}({\sigma_m } ; {0,{\gamma}_m^{ - 1}}) } $ \\
			\hline
			$f_F(\boldsymbol\gamma) \triangleq \prod_M G a\left(\gamma_m; \epsilon, \eta\right)$ \\
			\bottomrule
		\end{tabular}
	\end{table}
	
\begin{figure}
	\centering
	\includegraphics[width=1.02\linewidth]{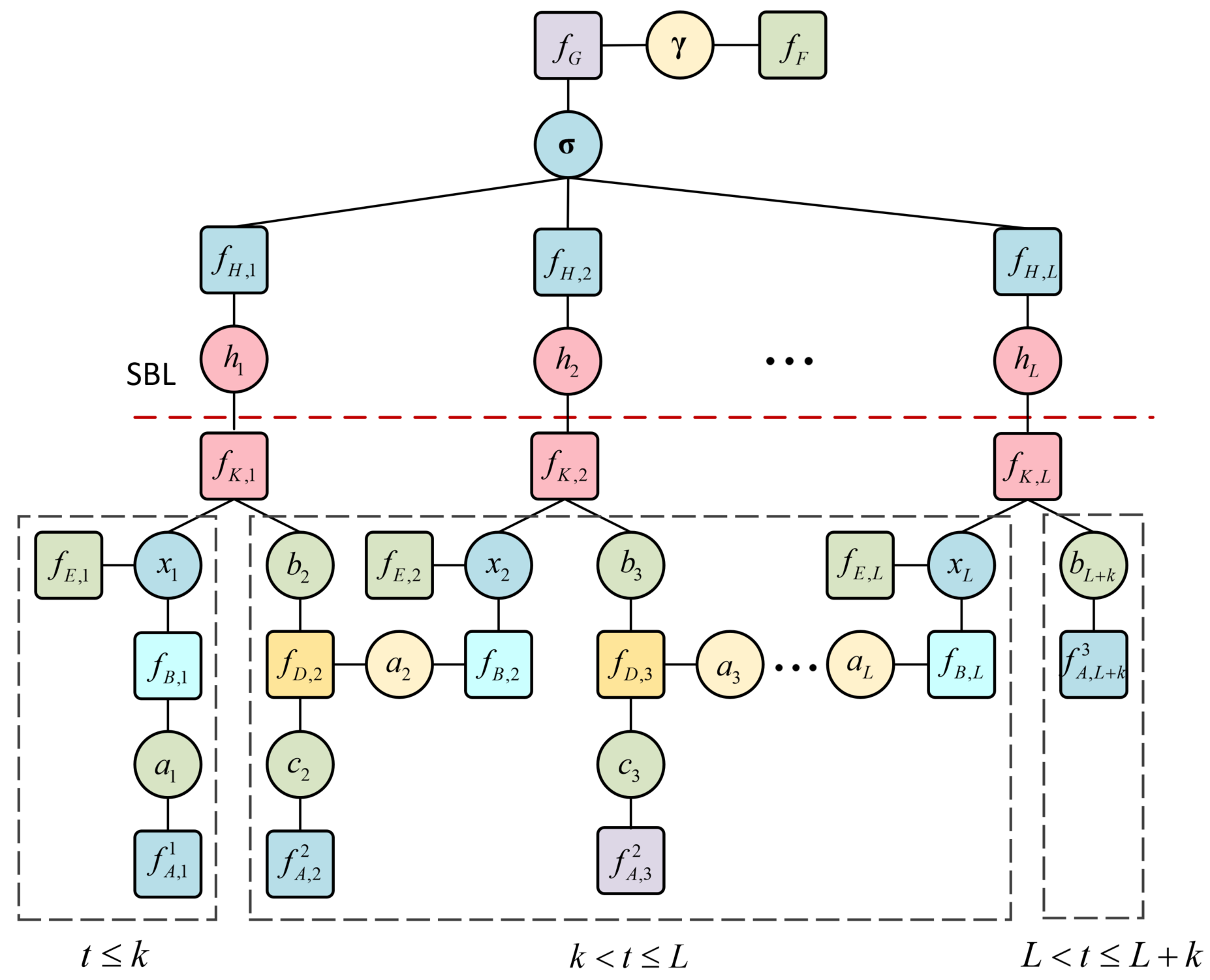}
	\caption{Factor graph representation for joint imaging and communication problem as formulated by \eqref{global_F}.}
	\label{factor_graph}
\end{figure}

Factor graphs and message passing are a powerful tool for inference and estimation. In this subsection, a factor graph model for \eqref{jt_distb} is established, based on which we can then jointly estimate $\mathbf{x}$ and ${\boldsymbol{\sigma}}$. To simplify the notations, we use $x_t$, $y_t$ and $\boldsymbol{\Theta}_t$ to represent $x(t)$, $y(t)$ and $\boldsymbol{\Theta}(t)$ respectively and we also define some variables and functions in Table I.  With these abbreviations and definitions, \eqref{piror}, \eqref{two_layer_prior}  and \eqref{pyt1} - \eqref{pyt3} can be respectively rewritten as
	
	\begin{equation} \label{px_nex}
	\begin{aligned} 
       p(\mathbf{x}) = \prod_{t=1}^{L} {f_{E,t}}\left(x_t \right),
	\end{aligned} 
   \end{equation} 	

	\begin{equation}\label{p_sig_nex}
		\begin{aligned} 
			p(\boldsymbol\sigma) 
			& = f_G(\boldsymbol{\sigma}, \boldsymbol\gamma) f_F(\boldsymbol\gamma),
		\end{aligned} 
	\end{equation}
	
	\begin{equation}\label{y_x1_nex}
		\begin{aligned} 
			p(y_t|x_t) & = p(y_t|a_t) p(a_t|x_t) \\
			& = f_{A,t}^1({a_t}) f_{B,t}({a_t},x_t),  t\leq k
		\end{aligned} 
	\end{equation}
	
	\begin{equation}\label{y_x2_nex}
		\begin{aligned} 
			& p\left( y_t|x_t, x_{t-k}, \boldsymbol{\sigma} \right) \\
			 = & \ p(y_t|c_t) p(c_t|a_t,b_t) p(a_t|x_t) p(b_t|x_{t-k},h_{t-k}) p(h_{t-k}|\boldsymbol\sigma)\\
			 = &\ f_{A,t}^2({c_t})f_{D,t}({a_t},{b_t},{c_t})f_{B,t} \left( {a_t},x_t \right)\\
			&\! \cdot f_{K,t-k}\left( {h_{t-k}},x_{t-k},{b_t} \right)  f_{H, t-k}(\boldsymbol{\sigma}, h_{t-k}), k< t\leq L  
		\end{aligned} 
	\end{equation}
	and
	\begin{equation}\label{y_x3_nex}
		\begin{aligned} 
			& p\left( y_t | x_{t-k}, \boldsymbol{\sigma} \right)  
			 = p \left( y_t|b_t \right)   p \left( b_t|x_{t-k},h_{t-k}\right)  p(h_{t-k}|\boldsymbol\sigma) \\
			& = f_{A,t}^3({b_t}) f_{K,t-k}\left( {h_{t-k}},x_{t-k},{b_t} \right)
			 f_{H, t-k}(\boldsymbol{\sigma}, h_{t-k}), \\
			& L < k \leq L+k.
		\end{aligned} 
	\end{equation}		
Then the joint distribution of $\mathbf{x}, \boldsymbol{\sigma}, a_t,b_t,c_t,h_t$ and $\boldsymbol{\gamma}$ conditioned on $\mathbf{y}$ can be given by
	\begin{equation} \label{global_F}
		\begin{aligned} 
			 &p(\mathbf{x}, \boldsymbol{\sigma}, a_t,b_t,c_t,h_t,\boldsymbol{\gamma} | \mathbf{y})\\
			 \propto \ & f_E(\mathbf{x}) f_G(\boldsymbol{\sigma}, \boldsymbol\gamma) f_F(\boldsymbol{\gamma})\cdot \\
			 &\cdot \prod_{t=1}^k f_{A,t}^1({a_t}) f_{B,t}({a_t},x_t) \\
			 &\cdot \prod_{t=k+1}^{L} f_{A,t}^2({c_t})f_{D,t}({a_t},{b_t},{c_t})f_{B,t}({a_t},x_t) \\ &
			 \qquad \quad \ \cdot f_{K,t-k}({h_{t-k}},x_{t-k},{b_t}) \cdot f_{H, t-k}(\boldsymbol{\sigma}, h_{t-k}) \\
			 &\cdot \prod_{t=L+1}^{L+k}  f_{A,t}^3({b_t}) f_{K,t-k}({h_{t-k}},x_{t-k},{b_t}) 
			 f_{H, t-k}(\boldsymbol{\sigma}, h_{t-k}).
		\end{aligned} 
	\end{equation}                                                      
Accordingly, the marginals of $\mathbf{x}$ and $\boldsymbol{\sigma}$ can be expressed by
\begin{equation} \label{marg_x}
	\mathcal{M}(\mathbf{x}) = \int_{\sim\{\mathbf{x}\}} p\left(\mathbf{x}, \boldsymbol{\sigma}, a_t, b_t, c_t, h_t, \boldsymbol{\gamma} \mid \mathbf{y}\right)
\end{equation}
and
\begin{equation}\label{marg_sigma}                 \mathcal{M}(\boldsymbol{\sigma}) = \int_{\sim\{\boldsymbol{\sigma}\}} p\left(\mathbf{x}, \boldsymbol{\sigma}, a_t, b_t, c_t, h_t, \boldsymbol{\gamma} \mid \mathbf{y}\right) .     
\end{equation}
Based on the factorization in \eqref{global_F}, we take $k=1$ as an example and establish the factor graph for joint communication and imaging as shown in Fig. \ref{factor_graph}, where we drop the augments of the functions for simplicity of notations. 

It is noteworthy that though $\mathcal{M}(\mathbf{x})$ and $\mathcal{M}(\boldsymbol{\sigma})$ are given in \eqref{marg_x} and \eqref{marg_sigma}, it is difficult to maximize them due to the high-dimensional integral. However, the factor graph in Fig.\ref{factor_graph} provides us a possible way to find approximation marginal functions of $\mathcal{M}(\mathbf{x})$ and $\mathcal{M}(\boldsymbol{\sigma})$, which are expected to allow easy maximization.

  \subsection{Message Passing Based Echoes Decoupling Algorithm}
		Considering that the factor graph contains loops, so iterative message passing is needed.  We divide the graph into upper and lower parts by the dotted line, where the lower part represents the message update for $\mathbf{x}$ and the upper part represents the message update for $\boldsymbol{\sigma}$. To estimate $\mathbf{x}$ and $\boldsymbol{\sigma}$ jointly, the upper and lower parts of Fig.\ref{factor_graph} 
		are updated alternatively, and the adaptive Sparse Bayesian learning (SBL) algorithm in \cite{luo2021unitary} is adopted to recover $\boldsymbol{\sigma}$. 
		
  Based on the sum-product algorithm (SPA) \cite{loeliger2004an,procdi_facgph}, the belief (or marginal function)  of $x_t$ can be calculated by the product of all incoming messages to  $x_t$, i.e.,
		\begin{equation} \label{b_xt}
			B (x_t) = \mu _{{f_{E,t}} \to {x_t}} (x_t) \cdot {\mu _{{f_{B,t}} \to {x_t}}} (x_t) \cdot  \mu _{{f_{K,t}} \to {x_t}} (x_t).
		\end{equation}
		where 
		\begin{equation}
		\mu _{{f_{E,t}} \to {x_t}} = \frac{1}{4}\sum_{i=1}^{4} \delta [x_t- e^{j(\frac{\pi}{2}i - \frac{\pi}{4})}].
		\end{equation}
To obtain $B (x_t)$, we first need to derive messages in Fig. \ref{factor_graph} from time index 1 to $L+k$ successively to update ${\mu _{{f_{B,t}} \to {x_t}}} (x_t)$. After that, we need to derive messages in reverse order of time index, i.e., form time index $L+k$ to $1$, to update $\mu _{{f_{K,t}} \to {x_t}} (x_t)$.
     
       
		
		
				
 \emph{ 1) When $t\leq k$:} Based on \eqref{recv_sig} and \eqref{pyt1}, we have
		\begin{equation}
			\begin{aligned}
				{\mu _{{f^1_{A,t}} \to {a_t}}}(a_t) & = \mathcal N \left( {{a_t}; {m_{{f^1_{A,t}} \to {a_t}}}, W_{{f^1_{A,t}} \to {a_t}}^{ - 1}} \right)\\
				& = \mathcal N \left( {a_t}; y_t, \xi^2 \right).
			\end{aligned}			
		\end{equation}
		According to the structure of the factor graph, the message $\mu _{{a_t} \to {f_{B,t}}}(a_t)$ is forward to $f_{B,t}$, i.e.,
		\begin{equation}
			\mu _{{a_t} \to {f_{B,t}}}(a_t) = {\mu _{{f^1_{A,t}} \to {a_t}}}(a_t) = \mathcal N \left( {a_t}; y_t, \xi^2 \right).
		\end{equation}
		Based on SPA, we have 
		\begin{equation}
			\begin{aligned}
			{\mu _{{f_{B,t}} \to {x_t}}} (x_t) &= \int {{f_{B,t}}} \left( {{a_t},{x_t}} \right) \cdot {\mu _{{a_t} \to {f_{B,t}}}}(a_t)  {\rm{d}}{a_t}.
	    	\end{aligned}
		\end{equation}
		Since ${\mu _{{a_t} \to {f_{B,t}}}}(a_t)$ is Gaussian message and $f_{B,t}\left( {{a_t},{x_t}} \right)$ indicates a linear relationship between $a_t$ and $x_t$, $\mu _{{f_{B,t}} \to {x_t}}(x_t)$ also has a Gaussian form, i.e.,
		\begin{equation}\label{fb_x1}
			\begin{aligned}
				{\mu _{{f_{B,t}} \to {x_t}}} (x_t) = \mathcal N \left( {{x_t}; {m_{{f_{B,t}} \to {x_t}}}, W_{{f_{B,t}} \to {x_t}}^{ - 1}} \right),
			\end{aligned}
		\end{equation}		
		where ${m_{{f_{B,t}} \to {x_t}}}$ and $W_{{f_{B,t}} \to {x_t}}$ are calculated by 
		\begin{equation}
			{W_{{f_{B,t}} \to {x_t}}} = {({{\alpha _c\bf{g}}^{\rm{T}}}{{\bf{\Theta }}_t}{{\bf{h}}_{\bf{c}}})^\mathrm H}{W_{{f^1_{A, t}} \to {a_t}}}({\alpha _c{\bf{g}}^{\rm{T}}}{{\bf{\Theta }}_t}{{\bf{h}}_{\bf{c}}}),
		\end{equation}
    	and
		\begin{equation}
			{m_{{f_{B,t}} \to {x_t}}} = W_{{f_{B,t}} \to {x_t}}^{ - 1}{({\alpha _c{\bf{g}}^{\rm{T}}}{{\bf{\Theta }}_t}{{\bf{h}}_{\bf{c}}})^\mathrm H}({W_{{f^1_{A,t}} \to {a_t}}}{m_{{f^1_{A,t}} \to {a_t}}}).
		\end{equation}
	The message outgoing from $x_t$ can be calculated by
	\begin{equation}\label{x_out}
	 \begin{aligned}
		&\mu_{x_t \to f_{K,t}}(x_t) = \mu_{f_{E,t} \to x_t}(x_t) \cdot \mu _{{f_{B,t}} \to {x_t}}(x_t) \propto                                                                       
		 \\ & \sum_{i=1}^{4} \delta (x_t- e^{j(\frac{\pi}{2}i - \frac{\pi}{4})}) \cdot \mathcal N \left( {{x_t}; {m_{{f_{B,t}} \to {x_t}}}, W_{{f_{B,t}} \to {x_t}}^{ - 1}} \right)
	 \end{aligned}
	\end{equation}
	Unfortunately, the message in \eqref{x_out} is no longer Gaussian, which makes the subsequent message calculation intractable. To this end, we approximate $\mu_{x_t \to f_{K,t}} (x_t)$ to a Gaussian message $\tilde{\mu}_{x_t \to f_{K,t}}= \mathcal N \left( {{x_t}; {m_{x_t \to f_{K,t}}}, V_{x_t \to f_{K,t}}} \right) $ by minimizing the Kullback-Leibler (KL) divergence between them. The optimal solution is given by moment matching, i.e.,  
	\begin{equation}\label{mom1}
		\begin{aligned}			
	     &m_{x_t \to f_{K,t}} = \mathbb{E}_{\mu_{x_t \to f_{K,t}}} (x_t) \\
	       & =  \frac{1}{4}\sum\limits_{i = 1}^4 {\exp \left( - {\left| {e^{j(\frac{\pi}{2}i - \frac{\pi}{4})} - {m_{{f_{B,t}} \to {x_t}}}}\right|^2}{W_{{f_{B,t}} \to {x_t}}}\right)} ,
	    \end{aligned}
    \end{equation}	
    and
	\begin{equation}\label{mom2}
		\begin{aligned}
	      &V_{x_t \to f_{K,t}} = \mathbb{E}_{\mu_{x_t \rightarrow f_{K, t}}}\left(\left|x_t-\mathbb{E}_{\mu_{x_t \rightarrow f_{K, t}}}\left(x_t\right)\right|^2\right) = \\
	      & \frac{1}{4} \sum_{i=1}^4\!\left[\exp \! \left(\!\!-\! \left|e^{j(\frac{\pi}{2}i - \frac{\pi}{4})}-m_{f_{B,t} \rightarrow x_t}\right|^2 \! W_{f_{B,t} \rightarrow x_t}\!\! \right)\!-\!m_{x_t \rightarrow f_{K,t}} \! \right]^2
	   \end{aligned}
    \end{equation}
		
   \emph{ 2) When $k<t\leq L$ :} 
  Suppose that the message from the variable node $h_{t-k}$ to the factor node $f_{K,{t-k}}$, i.e., $\mu_{h_{t-k} \to f_{K,{t-k}}}(h_{t-k})$ is given in the last round of iteration, with its belief denoted by $B(h_{t-k})$. According to the BP rule, the message outgoing from $f_{K,{t-k}}$ to $b_t$ can be expressed by
   \begin{equation} \label{fk_b}
   	\begin{aligned}
   		&\mu_{f_{K,{t-k}} \to b_{t} } (b_{t}) =  \int\!\!\!\int {{f_{K,{t-k}}}} \left( {{h_{t-k}},{x_{t-k}},b_{t}} \right) \\& \cdot {\mu _{{h_{t-k}} \to {f_{K,{t-k}}}}}(h_{t-k})  \cdot {\mu _{{x_{t-k}} \to {f_{K,{t-k}}}}}(x_{t-k}) {\rm{d}}{h_{t-k}}{\rm{d}}x_{t-k}.
   	\end{aligned}
   \end{equation}	
   It can be seen that the computation in \eqref{fk_b} is difficult due to the involved complex integral. Although numerical calculation seems feasible, it will suffer from high computation complexity and make the subsequent message update intractable due to the lack of closed-form expression. To simplify the message computation, we approximate the incoming message $\mu _{{h_{t-k}} \to {f_{K,{t-k}}}}(h_{t-k})$ to the Dirac delta function with proper location, i.e.,  
\begin{equation} \label{delta_ht_fK}
	\mu _{{h_{t-k}} \to {f_{K,{t-k}}}}(h_{t-k}) \approx \delta(h_{t-k} - m_{h_{t-k}}),
\end{equation}
where $B(h_{t-k})$ is maximized at $m_{h_{t-k}}$. Substituting \eqref{delta_ht_fK} into \eqref{fk_b} yields 
	\begin{equation} \label{fk_b2}
	\begin{aligned}
		&\mu_{f_{K,{t-k}} \to b_{t} } (b_{t}) \approx \tilde{\mu}_{f_{K,{t-k}} \to b_{t} } (b_{t}) \\= & \int {{f_{K,{t-k}}}} \left( {m_{h_{t-k}},{x_{t-k}},b_{t}} \right) \cdot {\mu _{{x_{t-k}} \to {f_{K,{t-k}}}}}(x_{t-k}) {\rm{d}}x_{t-k}.
	\end{aligned}
\end{equation}	                         
It is not hard to show that $\tilde{\mu}_{f_{K,{t-k}} \to b_{t} } (b_{t})$ still preserves Gaussianity, i.e.,
 \begin{equation}  \label{37}
 	\tilde{\mu}_{f_{K,{t-k}} \to b_{t} } (b_{t}) = \mathcal N \left( {{b_{t}}; {m_{{f_{K,{t-k}}} \to {b_{t}}}}, V_{{f_{K,{t-k}}} \to {b_{t}}}} \right),
 \end{equation}
where 
\begin{equation}
	V_{{f_{K,{t-k}}} \to {b_{t}}} = m_{h_{t-k}} V_{x_{t-k} \to f_{K,{t-k}}} (m_{h_{t-k}} )^\text{H} 
\end{equation}
and                    
\begin{equation}                                                           
	m_{{f_{K,{t-k}}} \to {b_{t}}} = m_{h_{t-k}} \cdot  m_{x_{t-k} \to f_{K,{t-k}}}. 
\end{equation}
It is clear that 
	\begin{equation} \label{miu_bt_fD}
		\begin{aligned}
			\mu_{b_t \to f_{D,t} }(b_t) = \mu_{f_{K,{t-k}} \to b_t }(b_t) . 
		\end{aligned}
	\end{equation}	
The message outgoing from the function node $f_{D,t}$ to the variable node $a_t$ can be given by
	\begin{equation} \label{fk_b3}
	\begin{aligned}
		&\mu_{f_{D,t} \to a_t} (a_t) = \\& \int\!\!\int {{f_{D,t}}} \left( {a_t,b_t,c_t} \right) \cdot {\mu _{{b_t} \to {f_{D,t}}}}(b_t) \cdot {\mu _{{c_t} \to {f_{D,t}}}}(c_t) {\rm{d}}{b_t}{\rm{d}}c_t
	\end{aligned}
\end{equation}	
where 
\begin{equation} \label{miu_c_fD}
	\begin{aligned}
	&{\mu _{{c_t} \to {f_{D,t}}}}(c_t) = {\mu _{{f^2_{A,t}} \to {c_t}}}(c_t) \\
	=&\mathcal N \left( {{c_t}; {m_{{f^2_{A,t}} \to {c_t}}}, V_{{f^2_{A,t}} \to {c_t}}} \right)
    = \mathcal N \left( {c_t}; y_t, \xi^2 \right)
\end{aligned} 
\end{equation}
With the combination of \eqref{miu_bt_fD} and  \eqref{miu_c_fD}, we have
\begin{equation}
\mu_{f_{D,t} \to a_t} (a_t) =\mathcal N \left( {a_t}; {m_{f_{D,t} \to a_t}, V_{f_{D,t} \to a_t}} \right), 
\end{equation}
where
\begin{equation}
	{m_{{f_{D,t}} \to {a_t}}} = {m_{{c_t} \to {f_{D,t}}}} - {m_{{b_t} \to {f_{D,t}}}},
\end{equation}

\begin{equation}
	{V_{{f_{D,t}} \to {a_t}}} = {V_{{c_t} \to {f_{D,t}}}} + {V_{{b_t} \to {f_{D,t}}}}.
\end{equation}
and

Based on the structure of the factor graph, we have 
\begin{equation}
	\mu_{a_t \to f_{B,t}} (a_t) = \mu_{f_{D,t} \to a_t} (a_t),
\end{equation}
and 
\begin{equation}\label{fb_x2}
	\begin{aligned}
		\mu_{f_{B,t} \to x_t} (x_t) &= \int f_{B,t}(a_t,x_t)\cdot \mu_{a_t \to f_{B,t}} (a_t) d a_t \\
		& =  \mathcal N \left( {{x_t}; {m_{{f_{B,t}} \to {x_t}}}, W_{{f_{B,t}} \to {x_t}}^{ - 1}} \right)
	\end{aligned}
\end{equation}
where
\begin{equation}
	{W_{{f_{B t}} \to {x_t}}} = {({{\alpha _c\bf{g}}^{\rm{T}}}{{\bf{\Theta }}_t}{{\bf{h}}_{\bf{c}}})^{\rm{H}}}{V_{{f_{D,t}} \to {a_t}}^{-1}}({\alpha _c{\bf{g}}^{\rm{T}}}{{\bf{\Theta }}_t}{{\bf{h}}_{\bf{c}}}),
\end{equation}
and
\begin{equation}
	{m_{{f_{B,t}} \to {x_t}}} = W_{{f_{B,t}} - {x_t}}^{ - 1}{({\alpha _c{\bf{g}}^{\rm{T}}}{{\bf{\Theta }}_t}{{\bf{h}}_{\bf{c}}})^{\text H}}({V_{{f_{D,t}} \to {a_t}}^{-1}}{m_{{f_{D,t}} \to {a_t}}}).
\end{equation}                                                                                       
Note that the non-Gaussian incoming message to $x_t$ will make the outgoing message $\mu_{x_t \to f_{K,t}}(x_t)$ no longer keeps the Gaussian form. We here approximate $\mu_{x_t \to f_{K,t}}(x_t)$ to a Gaussian form based on moment matching with the same way in \eqref{x_out} - \eqref{mom2}, i.e.,
\begin{equation} \label{miu_x_fK_gauss}
	\begin{aligned}
	&\mu_{x_t \to f_{K,t}}(x_t)= \mu_{f_{E,t} \to x_t}(x_t) \cdot \mu _{{f_{B,t}} \to {x_t}}(x_t) \\
	& \propto \sum_{i=1}^{4} \delta \left[x_t- e^{j(\frac{\pi}{2}i - \frac{\pi}{4})}\right] \cdot \mathcal N \left( {{x_t}; {m_{{f_{B,t}} \to {x_t}}}, W_{{f_{B,t}} \to {x_t}}^{ - 1}} \right) \\
	&\approx \tilde{\mu}_{x_t \to f_{K,t}}(x_t) = \mathcal N \left( {{x_t}; {m_{x_t \to f_{K,t}}}, V_{x_t \to f_{K,t}}} \right),	
	 \end{aligned}
\end{equation}
where $m_{x_t \to f_{K,t}}$ and $V_{x_t \to f_{K,t}}$ have same expressions as \eqref{mom1} and \eqref{mom2}.

We have derived $\mu_{x_t \to f_{K,t}}(x_t)$ for $t\leq k$ and $k <t\leq L$. We then need to derive $\mu _{{f_{K,t}} \to {x_t}} (x_t)$ in different time segments.
When $L < t \leq L+k $, based on \eqref{recv_sig} and \eqref{pyt3}, we have
\begin{equation} \label{miu_b_fK}
	\begin{aligned}
		&\mu_{b_{t} \to f_{K,t-k}}(b_{t}) = \mu_{f^3_{A,t} \to b_{t}}(b_{t}) \\ &=  \mathcal N \left( {{b_{t}}; {m_{{f^3_{A,t}} \to b_{t}}}, W_{{f^3_{A,t}} \to b_{t+k}}^{ - 1}} \right)\\
		& = \mathcal N \left( b_{t}; y_{t}, \xi^2 \right).
	\end{aligned}
\end{equation}

When $L-k < t \leq L$, \eqref{miu_b_fK} can be recast as 
$\mu_{b_{t+k} \to f_{K,t}}(b_{t+k}) = \mathcal N \left( b_{t+k}; y_{t+k}, \xi^2 \right) $. With the combination of the same approximation in \eqref{delta_ht_fK}, we have
\begin{equation} \label{fk_x}
	\begin{aligned}
      \mu_{f_{K,t} \to x_t}(x_t) = &\int\!\!\int {{f_{K,t}}} \left( {{h_t},{x_t},b_{t+k}} \right)\cdot {\mu _{{h_t} \to {f_{K,t}}}}(h_t) \\
       &  \cdot \mu_{b_{t+k} \to f_{K,t} }(b_{t+k}) {\rm{d}}{h_t}{\rm{d}}x_t \\
       \approx \!\int {{f_{K,t}}} &\left({{m_{h_t}},{x_t},b_{t+k}} \right) \cdot \mu_{b_{t+k} \to f_{K,t} }(b_{t+k}) {\rm{d}}x_t\\ 
	  =\mathcal N &\left( {{x_t}; {m_{{f_{K,t}} \to {x_t}}}, W_{{f_{K,t}} \to {x_t}}^{ - 1}} \right).
\end{aligned}      
\end{equation}
where 
\begin{equation}
m_{{f_{K,t}} \to {x_t}} = y_{t+k} / m_{h_t},
\end{equation}
\begin{equation}
	W_{{f_{K,t}} \to {x_t}} =\frac{m^{\rm{H}}_{{f_{K,t}} \to {x_t}}m_{{f_{K,t}} \to {x_t}}}{\xi^2}.
\end{equation}

When $k< t\leq L-k$, the message $\mu_{x_t \to f_{B,t}}(x_t)$ can be calculated by 
\begin{equation} \label{}
	\begin{aligned}
	&\mu_{x_t \to f_{B,t}}(x_t) = \mu_{f_{E,t} \to x_t}(x_t) \cdot \mu _{{f_{K,t}} \to {x_t}}(x_t)
	\\ &\propto \sum_{i=1}^{4} \delta \left(x_t- e^{j(\frac{\pi}{2}i - \frac{\pi}{4})}\right) \cdot \mathcal N \left( {{x_t}; {m_{{f_{K,t}} \to {x_t}}}, W_{{f_{K,t}} \to {x_t}}^{ - 1}} \right) \\
	&\approx \mathcal N \left( {{x_t}; {m_{x_t \to f_{B,t}}}, V_{x_t \to f_{B,t}}} \right).
	\end{aligned}
\end{equation}
where
	\begin{equation}\label{mom21}
	\begin{aligned}			
		&m_{x_t \to f_{B,t}} = \mathbb{E}_{\mu_{x_t \to f_{B,t}}} (x_t) \\
		& =  \frac{1}{4}\sum\limits_{i = 1}^4 {\exp \left( - {\left| {e^{j(\frac{\pi}{2}i - \frac{\pi}{4})} - {m_{{f_{K,t}} \to {x_t}}}}\right|^2}{W_{{f_{K,t}} \to {x_t}}}\right)} ,
	\end{aligned}
\end{equation}	
and
\begin{equation}\label{mom22}
	\begin{aligned}
		&V_{x_t \to f_{B,t}} = \mathbb{E}_{\mu_{x_t \rightarrow f_{B, t}}}\left(\left|x_t-\mathbb{E}_{\mu_{x_t \rightarrow f_{B, t}}}\left(x_t\right)\right|^2\right) = \\
		& \frac{1}{4} \sum_{i=1}^4\!\left[\exp \! \left(\!\!-\left|e^{j(\frac{\pi}{2}i - \frac{\pi}{4})}\!-\!m_{f_{K,t} \rightarrow x_t}\right|^2 \! W_{f_{K,t} \rightarrow x_t}\! \right)\!-\!m_{x_t \rightarrow f_{B,t}} \! \right]^2
	\end{aligned}
\end{equation}

Note that the same approximation method in \eqref{x_out} - \eqref{mom2} is adopted here to approximate $\mu_{x_t \to f_{B,t}}(x_t)$ to a Gaussian form. Then $\mu_{f_{B,t} \to a_t}(a_t)$ can be calculated by BP, i.e., 
\begin{equation}
\begin{aligned}
	&\mu_{f_{B,t} \to a_t}(a_t) = \int f_{B,t}\left(a_t,x_t \right) \mu_{x_t \to f_{B,t}}(x_t) d x_t   \\
 &=\mathcal N \left( {{a_t}; {m_{f_{B,t} \to a_t}}, V_{f_{B,t} \to a_t}} \right).
\end{aligned}
\end{equation}
where 
\begin{equation}
	{m_{{f_{B,t}} \to {a_t}}} = ({\alpha _c}{{\bf{g}}^{\rm{T}}}{{\bf{\Theta }}_t}{{\bf{h}}_{\bf{c}}}){m_{{x_t} \to {f_{B,t}}}}
\end{equation}
and
\begin{equation}
	{V_{{f_{B,t}} \to {a_t}}} = ({\alpha _c}{{\bf{g}}^{\rm{T}}}{{\bf{\Theta }}_t}{{\bf{h}}_{\bf{c}}}){V_{{x_t} \to {f_{B,t}}}}{({\alpha _c}{{\bf{g}}^{\rm{T}}}{{\bf{\Theta }}_t}{{\bf{h}}_{\bf{c}}})^{\rm{H}}}
\end{equation}
Based on SPA, we have $\mu_{a_t \to f_{D,t}}(a_t)  = \mu_{f_{B,t} \to a_t}(a_t)$ and 
\begin{equation} \label{fd_b0}
	\begin{aligned}
		&\mu_{f_{D,t} \to b_t} (b_t) = \\& \int\!\!\int {{f_{D,t}}} \left( {a_t,b_t,c_t} \right) \cdot {\mu _{{a_t} \to {f_{D,t}}}}(a_t) \cdot {\mu _{{c_t} \to {f_{D,t}}}}(c_t) {\rm{d}}{b_t}{\rm{d}}c_t\\
		&= \mathcal N \left( {{b_t}; {m_{f_{D,t} \to b_t}}, V_{f_{D,t} \to b_t}} \right)		
	\end{aligned}
\end{equation}	
and
\begin{equation}
  \mu_{b_t \to f_{K,t-k}} (b_t)= \mu_{f_{D,t} \to b_t} (b_t) .
\end{equation}
It can be verified that
   \begin{equation} \label{fk_x_last}
	\begin{aligned}
		&\mu_{f_{K,t-k} \to x_{t-k} } (x_{t-k}) =  \int\!\!\!\int {{f_{K,{t-k}}}} \left( {{h_{t-k}},{x_{t-k}},b_t} \right) \\ &\cdot {\mu _{{h_{t-k}} \to {f_{K,{t-k}}}}}(h_{t-k}) \cdot \mu_{b_t \to f_{K,t-k}} (b_t) {\rm{d}}{h_{t-k}}{\rm{d}}x_{t-k}.   
	\end{aligned}
\end{equation}	

\begin{algorithm}[t]
	\caption{Message Passing Based Joint Communication and Imaging Algorithm.}
	\begin{algorithmic}[1]
      \STATE  Initialize the maximum number of iterations $K_{max} $ , and threshold $\epsilon_0 $
      \WHILE {$\Delta \leq \epsilon_0$ and $p<K_{max}$  }
            \STATE For $t \leq k$, compute ${\mu _{{f_{B,t}} \to {x_t}}} (x_t)$ with \eqref{fb_x1}.
            \STATE For $ k<t \leq L$, compute $\mu_{f_{B,t} \to x_t} (x_t)$ with \eqref{fb_x2}.
            \STATE For $ L-k<t \leq L$, compute $\mu_{f_{K,t} \to x_t}(x_t)$ with \eqref{fk_x}.
            \STATE For $ k<t \leq L-k$, compute $\mu_{f_{K,t} \to x_{t} } (x_{t})$ with \eqref{appx_fk_x_last}.
            \STATE  Compute $B (x_t)$ with \eqref{b_xt2}.
            \STATE The estimate of $x_t$ in the $p$th iteration is given by $ x_t^p = \hat{x}_t$, where $\hat{x}_t^p$ maximizes $B (x_t)$.
            \STATE Compute $\boldsymbol\sigma^p$ by iterating \eqref{SBL1}- \eqref{SBL4} until convergence.
            \STATE Compute $\Delta =  \left\| \mathbf{x}^p - \mathbf{x}^{p-1} \right\| + \left\| \boldsymbol\sigma^p - \boldsymbol\sigma^{p-1}\right\| $.
       \ENDWHILE 
	\end{algorithmic}
	\label{Alg_mp}
\end{algorithm}

Substituting \eqref{delta_ht_fK} into \eqref{fk_x_last}, we have
\begin{equation}\label{appx_fk_x_last}
    \begin{aligned}
             & \mu_{f_{K,t-k} \to x_{t-k} } (x_{t-k})\approx \int\!\!\!\int {{f_{K,{t-k}}}} \left( {{h_{t-k}},{x_{t-k}},b_t} \right) \\ &\cdot {\mu _{{h_{t-k}} \to {f_{K,{t-k}}}}}(h_{t-k}) \cdot \delta(h_{t-k} - m_{h_{t-k}}) {\rm{d}}{h_{t-k}}{\rm{d}}x_{t-k}\\
             & = \int {{f_{K,{t-k}}}} \left( {m_{h_{t-k}},{x_{t-k}},b_t} \right) \cdot {\mu _{{h_{t-k}} \to {f_{K,{t-k}}}}}(h_{t-k}) \cdot  {\rm{d}}x_{t-k} \\
             & = \mathcal N \left( {x_{t-k}}; m_{{f_{K,{t-k}}} \to {x_{t-k}}}, W_{{f_{K,{t-k}}} \to x_{t-k}}^{ - 1} \right)
    \end{aligned}
\end{equation}
where 
\begin{equation}
    m_{{f_{K,{t-k}}} \to {x_{t-k}}} = {m_{f_{D,t} \to b_t}} / m_{h_{t-k}}
\end{equation}

\begin{equation}
    W_{{f_{K,{t-k}}} \to x_{t-k}} = m^{\rm{H}}_{{f_{K,{t-k}}} \to {x_{t-k}}}{W_{f_{D,t} \to b_t}}m_{{f_{K,{t-k}}} \to {x_{t-k}}}
\end{equation}

Based on \eqref{fb_x1}, \eqref{fb_x2}, \eqref{fk_x}, and \eqref{appx_fk_x_last},  $B (x_t)$ in \eqref{b_xt} can be calculated by

\begin{equation} \label{b_xt2}
	\begin{aligned}
	B (x_t)& = \frac{1}{4}\sum_{i=1}^{4} \delta [ x_t- e^{j(\frac{\pi}{2}i - \frac{\pi}{4})}] \\& \cdot \mathcal N \left( {{x_t}; {m_{{f_{B,t}} \to {x_t}}}, W_{{f_{B,t}} \to {x_t}}^{ - 1}} \right) \\& \cdot \mathcal N \left( {{x_t}; {m_{{f_{K,t}} \to {x_t}}}, W_{{f_{K,t}} \to {x_t}}^{ - 1}} \right)
	\end{aligned}
\end{equation}
The estimate of $x_t$ is given by $\hat{x}_t$, which maximize $B (x_t)$.  With the estimate of $\mathbf{x}$, the adaptive SBL algorithm \cite{luo2021unitary} can be adopted to update ${\boldsymbol\sigma}$. We define  
\begin{equation}
	\begin{small}
 	\mathbf{h} = \left[ y_{k+1}\!-\!\hat{a}_{{k}\!+\!1}, \cdots ,y_{L}\!-\!\hat{a}_{{L}},\ y_{L+1}, \cdots, y_{L+k}  \right]^{\rm{T}}.
 \end{small}
\end{equation}
where $\hat{a_t}={\alpha _c}{{\bf{g}}^{\rm{T}}}{\bf{\Theta }}(t){{\bf{h}}_{\bf{c}}}\hat{x}_t$. Then ${\boldsymbol\sigma}$ can be updated by iterating the following steps until ${{\boldsymbol\sigma}}$ converges.

\begin{equation} \label{SBL1}
	{{\bf{W}}_{\boldsymbol\sigma}} = { \frac{ {\bf{G}}^{\text H}{\bf{G}}}{\hat \xi} + {\mathop{\rm diag}\nolimits} (\hat{\bm{\gamma}} ) },
\end{equation}
\begin{equation}\label{SBL2}
	{{\boldsymbol\sigma}} =  \frac{ {{\bf{W}}_{\boldsymbol\sigma}}{{\bf{G}}^{\text H}}{\bf{h}}} {\hat \xi} ,
\end{equation}
%
\begin{equation} \label{SBL3}
	{\hat \gamma _m} = (2\epsilon + 1)/\left( {2\zeta  + {{\left| {{{\boldsymbol\sigma}}(m)} \right|}^2} + {{\bf{W}}^{-1}_{\boldsymbol\sigma}}(m, m)} \right),m = 1, \ldots , M
\end{equation}

\begin{equation} \label{SBL_add}
   \epsilon = \frac{1}{2} \sqrt{ \text{log} ( \frac{1}{M} \sum{\hat \gamma _m}) -\frac{1}{M}\sum{\text{log}(\hat \gamma _m) }                  }
\end{equation}

\begin{equation} \label{SBL4}
	\hat \xi  = \frac{{{{\left\| {{\bf{h}} - {\bf{G \boldsymbol\sigma }}} \right\|}^2}}}{{L - \sum\limits_{m=1}^M {{{\hat \gamma }_m}} }}
\end{equation}

\begin{figure}
	\centering
	\includegraphics[width=1\linewidth]{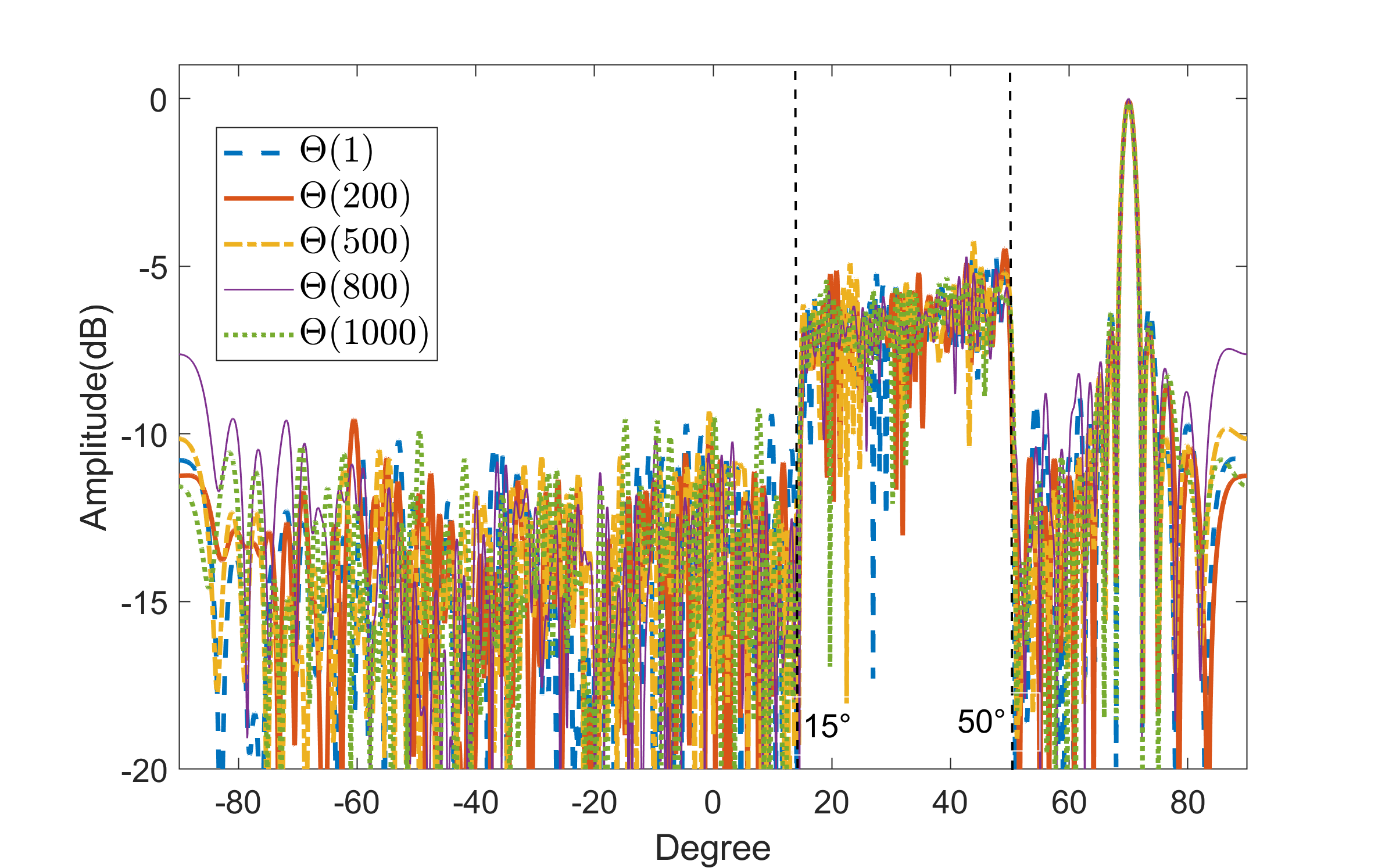}
	\caption{The receive beam pattern of the BS, i.e., ${{\bf{g}}^{\rm{T}}}{\bf{\Theta }}(t) a(\theta_p) $, in the case of different time index $t$.}
	\label{fig:fangxaingtu_randn}
\end{figure}

\begin{figure}
	\centering
	\includegraphics[width=1\linewidth]{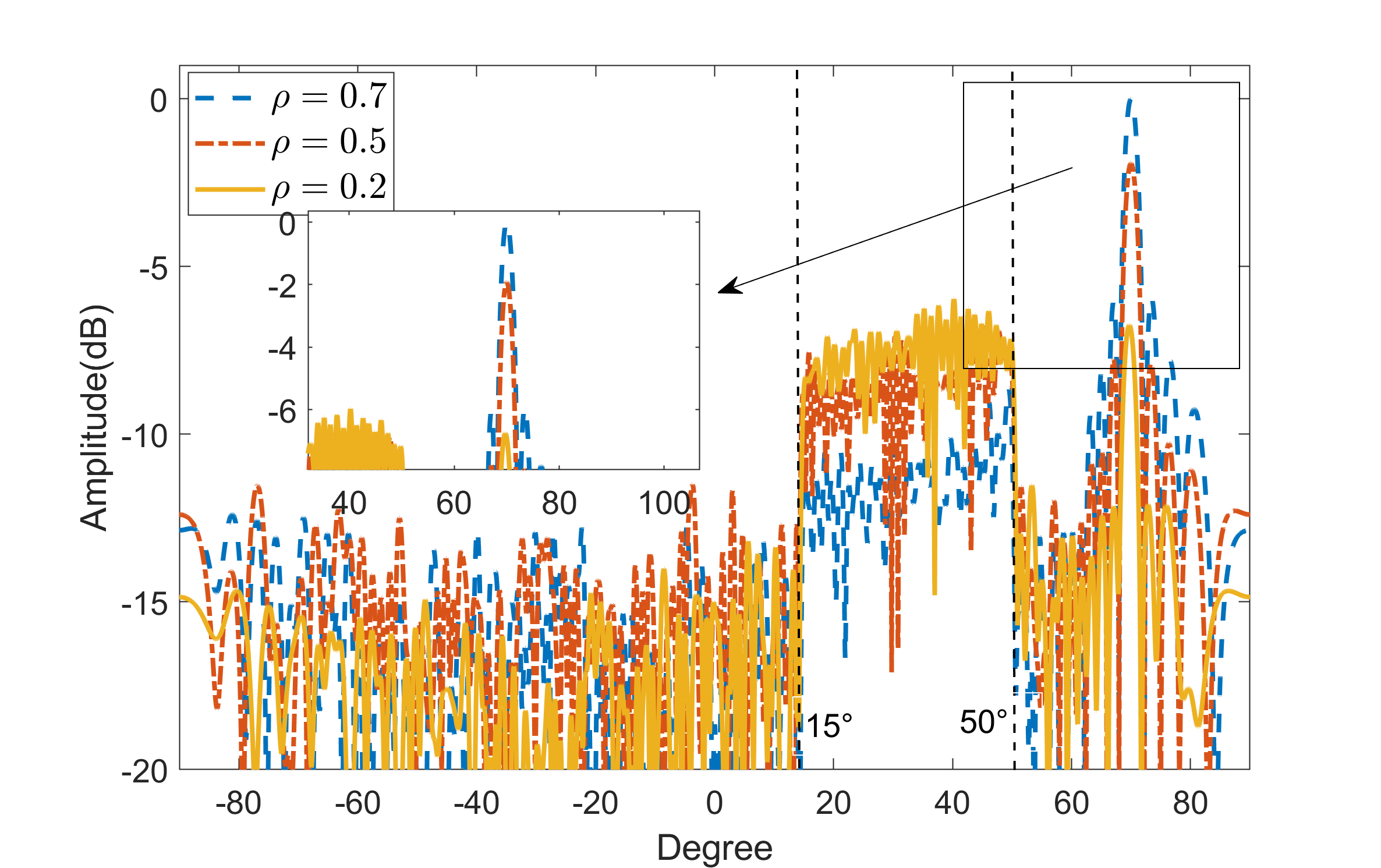}
	\caption{The receive beam pattern of the BS, i.e., ${{\bf{g}}^{\rm{T}}}{\bf{\Theta }}(t) a(\theta_p) $, in the case of different weight factor $\rho$. }
	\label{fig:fangxaingtu_rou}
\end{figure}

\begin{figure}
	\centering
	\includegraphics[width=1\linewidth]{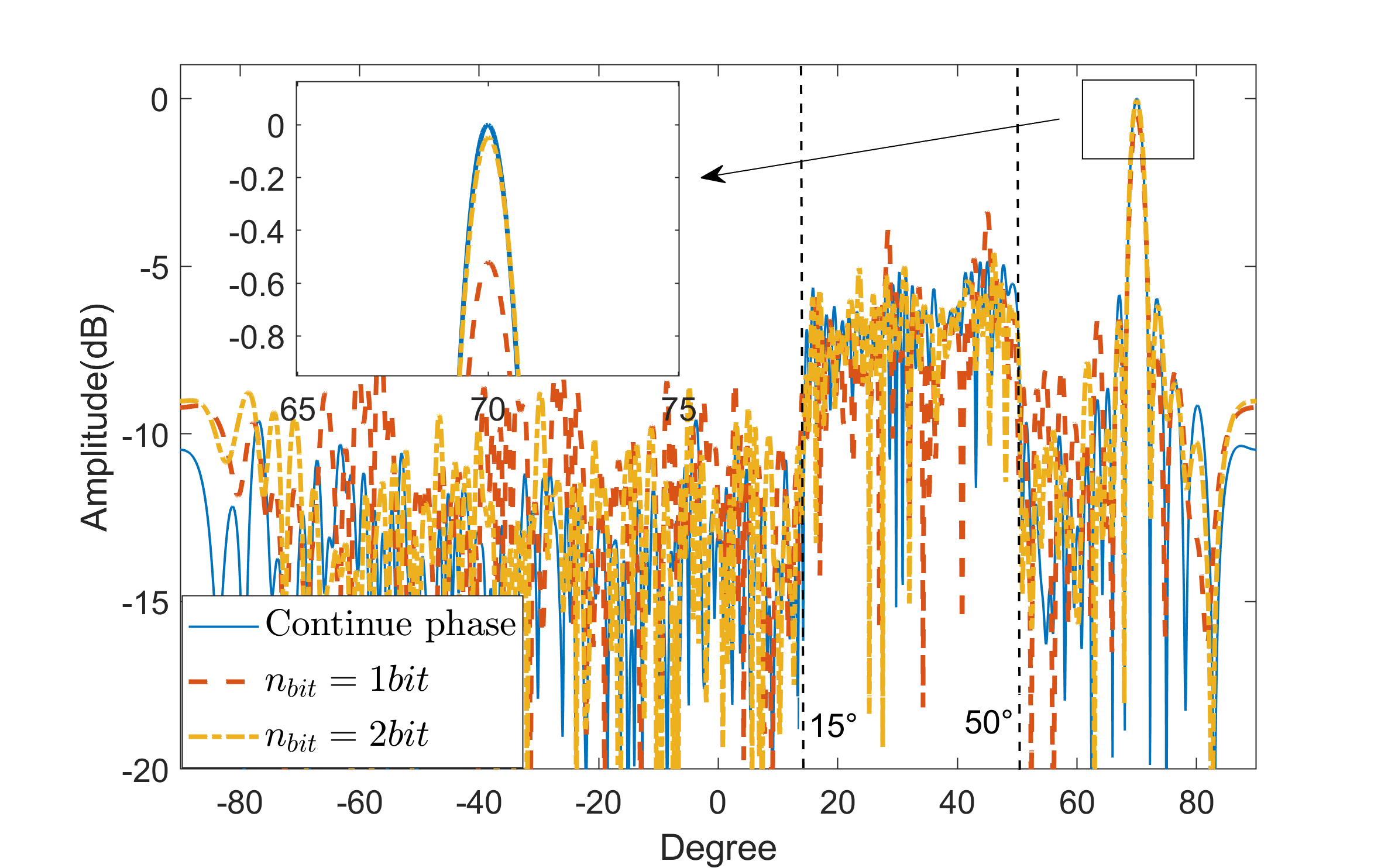}
	\caption{The receive beam pattern of the BS, i.e., ${{\bf{g}}^{\rm{T}}}{\bf{\Theta }}(t) a(\theta_p) $, in the case of different $n_{bit}$. }
	\label{fig:fangxaingtu_bit}
\end{figure}

		\begin{figure*}
			\centering
			\subfigure[The ground truth of $\boldsymbol{\sigma}$ in scenario 1.]{
				\includegraphics[height=0.16\linewidth]{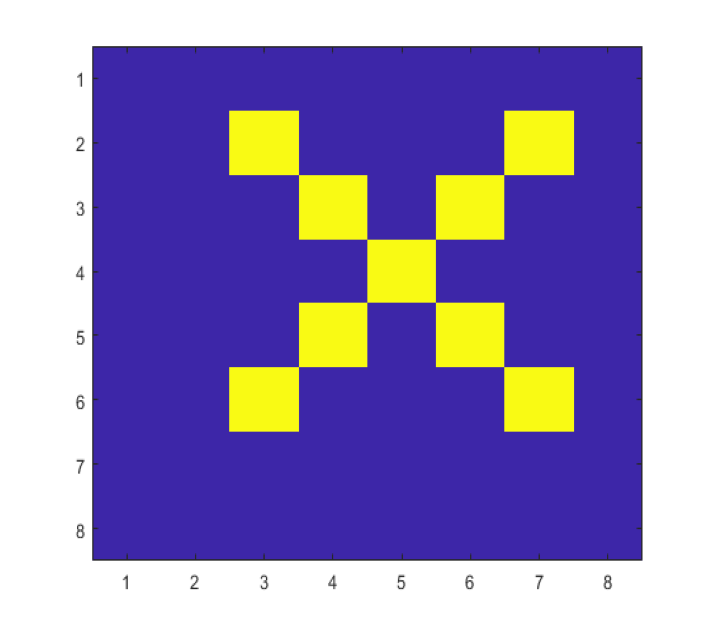}
			}
			\quad
			\subfigure[NMSE = -15.16 dB.]{
				\includegraphics[height=0.16\linewidth]{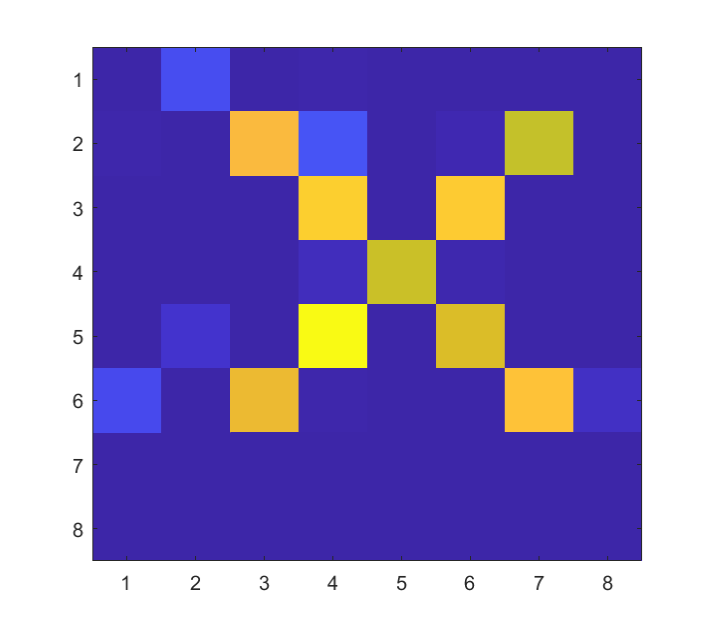}
			}
			\quad
			\subfigure[NMSE = -20.49 dB]{
				\includegraphics[height=0.16\linewidth]{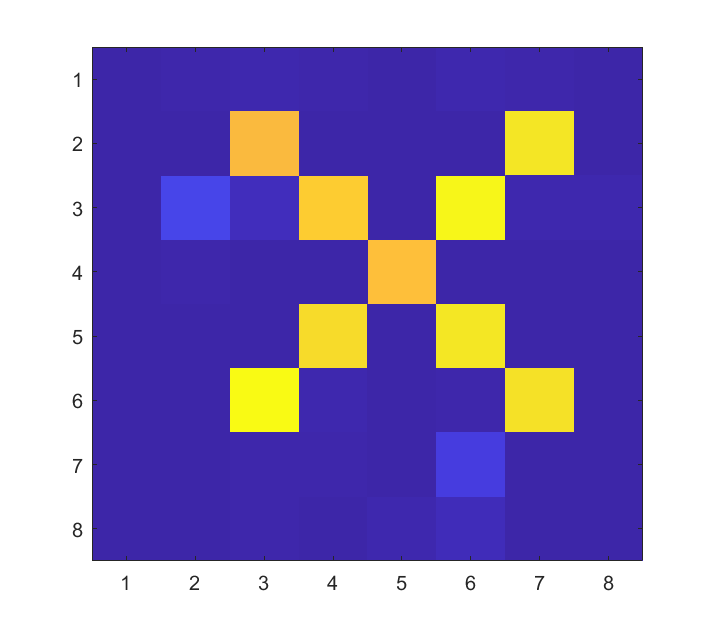}
			}
			\quad
			\subfigure[NMSE = -25.34 dB.]{
				\includegraphics[height=0.16\linewidth]{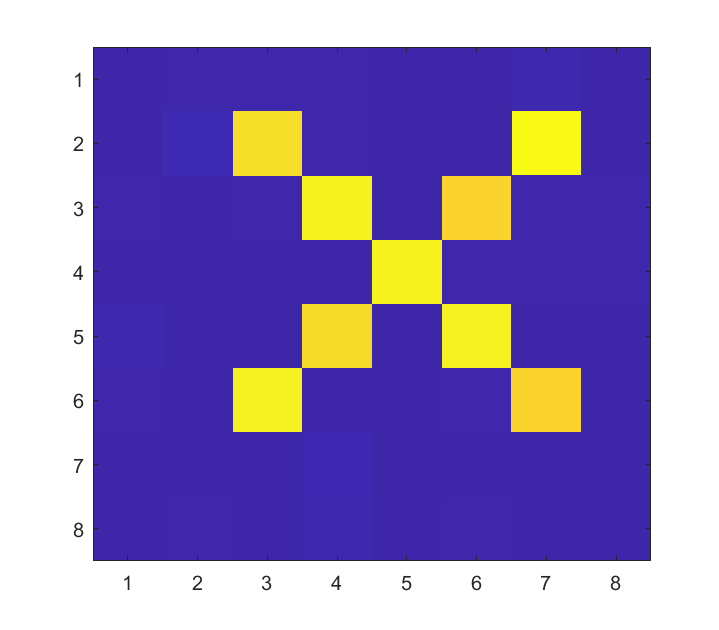}
			}
			\quad
			\subfigure[The ground truth of $\boldsymbol{\sigma}$ in scenario 2.]{
				\includegraphics[height=0.16\linewidth]{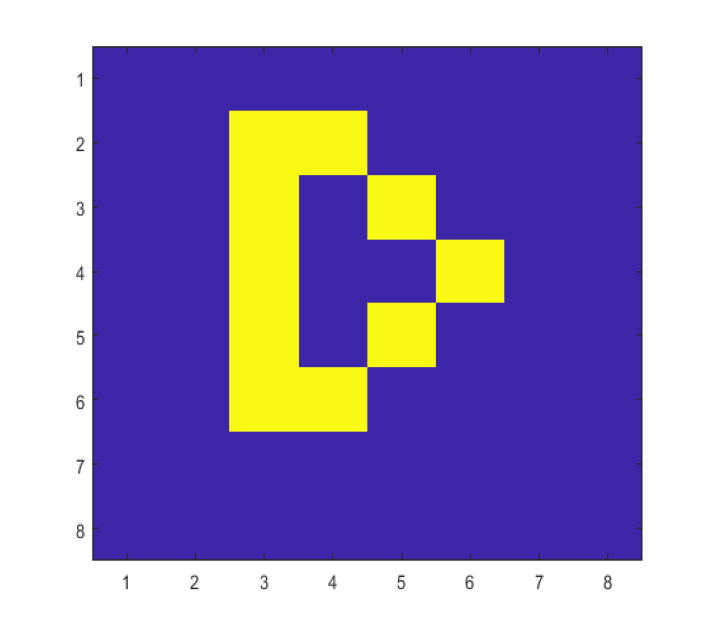}
			}
			\quad
			\subfigure[NMSE = -15.14 dB.]{
				\includegraphics[height=0.16\linewidth]{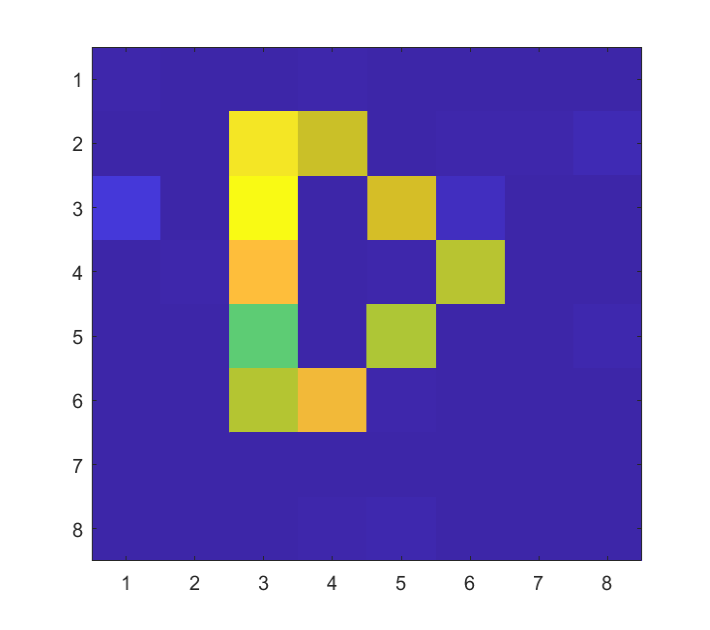}
			}
			\quad
				\subfigure[NMSE = -20.42 dB.]{
				\includegraphics[height=0.16\linewidth]{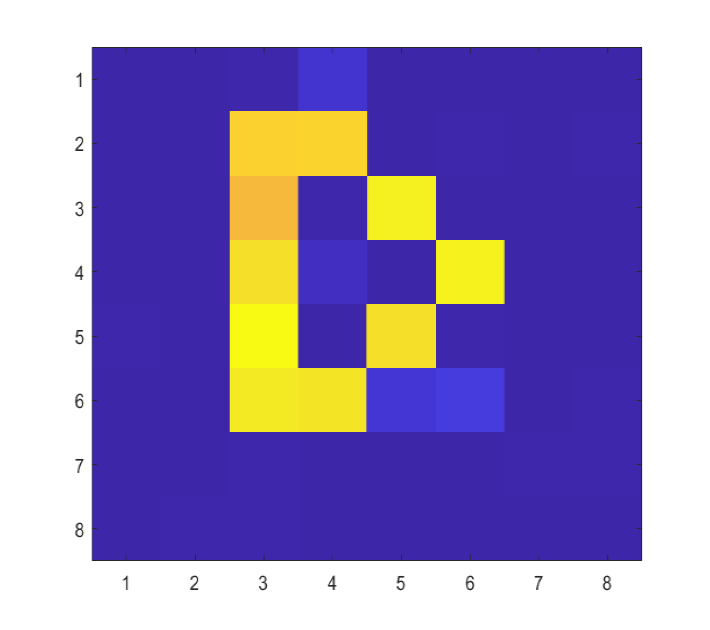}
			}
			\quad
				\subfigure[NMSE = -25.03 dB.]{
				\includegraphics[height=0.16\linewidth]{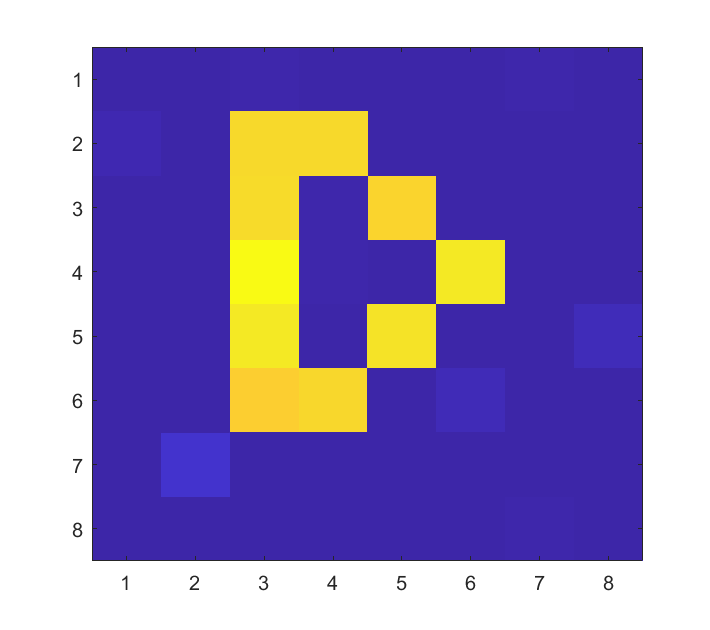}
			}
			\quad
				\subfigure[The ground truth of $\boldsymbol{\sigma}$ in scenario 3.]{
				\includegraphics[height=0.16\linewidth]{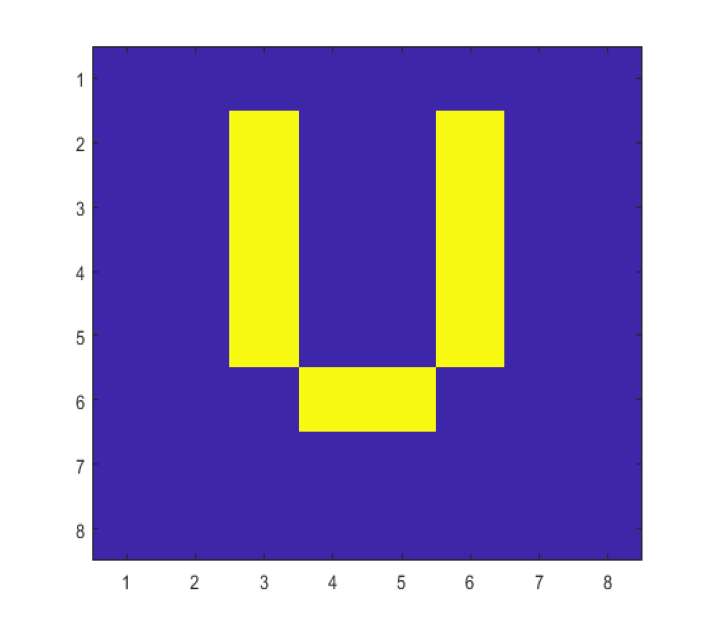}
			}
			\quad
			\subfigure[NMSE = -15.24 dB.]{
				\includegraphics[height=0.16\linewidth]{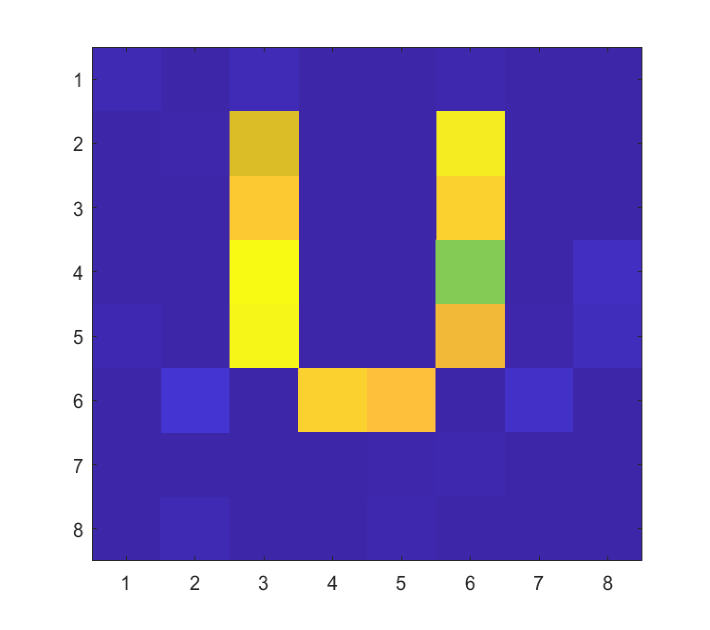}
			}
			\quad
			\subfigure[NMSE = -20.11 dB.]{
				\includegraphics[height=0.16\linewidth]{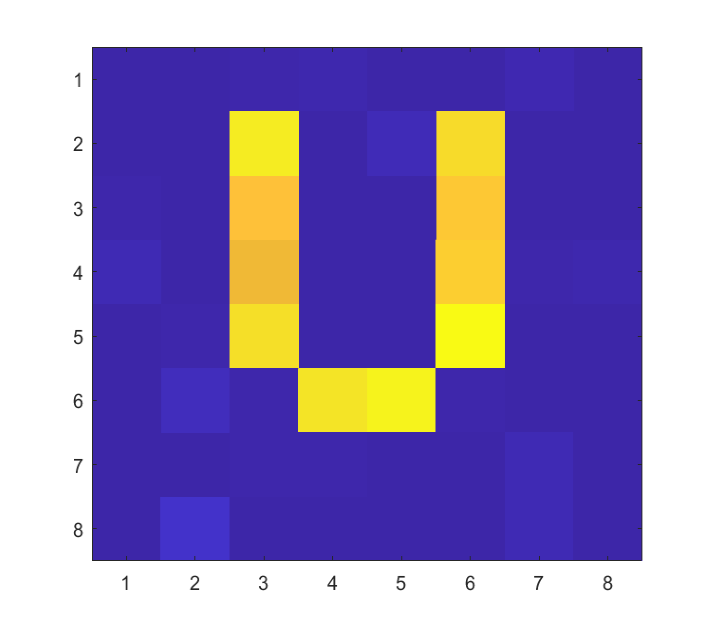}
			}
			\quad
			\subfigure[NMSE = -25.10 dB.]{
				\includegraphics[height=0.16\linewidth]{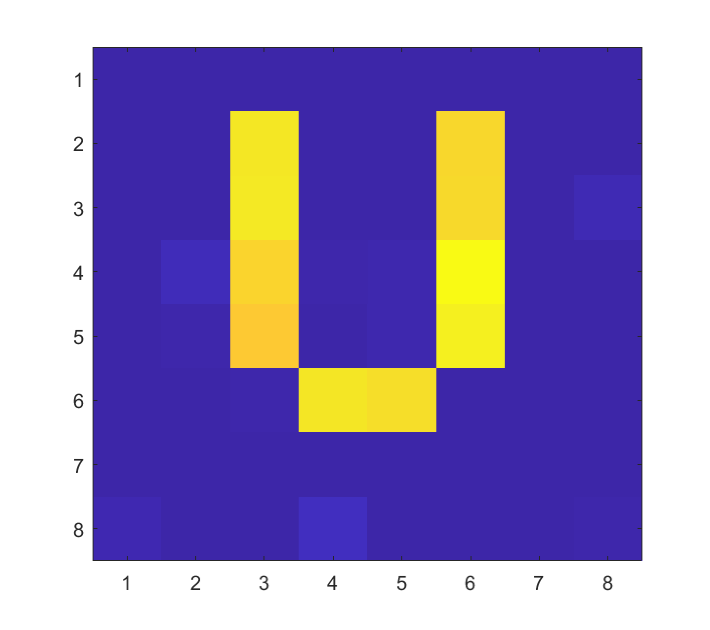}
			}
			\quad
			\caption{The ground truth of rearranged $\boldsymbol{\sigma}$ and corresponding imaging results under different NMSE levels.}	
			\label{XDU}
		\end{figure*}
The proposed message passing based joint communication and imaging algorithm is summarized in Algorithm 2.

	\section{Numerical Results}
     In this section, numerical experiments are conducted to evaluate the uplink ISAC system and the performance of proposed method. We consider a RIS with 150 elements and a BS located at ${\theta _B} = {-45^\circ}$ from the RIS. A single UE is located at ${\theta _U} = {70^\circ}$, and the transmitted signal is QPSK modulated with the length $L=1024$. The RoI is equidistantly divided into $M = 64$ grids, spacing from $15^ \circ$ to ${50^ \circ }$ from the RIS. In Algorithm 1, we set $\eta=0.001$ and $r=0.005$.

We define the communication-to-noise ratio (CNR) and the imaging-to-noise ratio (INR) as  
	\begin{equation}
		CNR = 10\text{log}_{10} \frac{\mathbb{E}_t \left[ \left| {\alpha_c}{{\bf{g}}^{\rm{T}}}{\bf{\Theta }}(t){{\bf{h}}_{\bf{c}}}x(t) \right|^2  \right] }{\xi^2}
	\end{equation}
	and 
	\begin{equation}
		INR = 10\text{log}_{10} \frac{\mathbb{E}_t \left[ \left| {{\alpha_I}{{\bf{g}}^{\rm{T}}}{\bf{\Theta }}(t){{\bf{H}}_{\bf{I}}}{\boldsymbol{\sigma }}x(t - k)} \right|^2  \right] }{\xi^2}.
	\end{equation}
	
	The performance of communication and imaging is assessed by the symbol-error-rate (SER) and normalized mean squared error (NMSE), which are respectively defined as 
		\begin{equation}
		SER = \sum_{q=1}^Q \frac{ E_q }{QL}
	\end{equation}
	and 
	\begin{equation}
		NMSE = 10\text{log}_{10} \frac{\Vert \boldsymbol{\sigma_q} -\boldsymbol{\sigma} \Vert ^2 }{\Vert \boldsymbol{\sigma} \Vert^2} ,
	\end{equation}
where $Q$ is the number of trials, $E_q$ and $\boldsymbol{\sigma}_q$ denote the number of error symbol and the estimate of $\boldsymbol{\sigma}$ for the $q$th trial, respectively.

\subsection{Phase Optimization Result}
 We first illustrate the performance of the proposed phase design scheme. Firstly, we set $\rho = 0.5$ and investigate the receive beam pattern of the BS at different time index $t$, i.e., ${{\bf{g}}^{\rm{T}}}{\bf{\Theta }}(t)a(\theta_p), \theta_p \in (-90^ \circ, 90^\circ)$. In Fig.\ref{fig:fangxaingtu_randn}, we show the receive beam pattern at time indexes $t=1,t=200,t=500,t=800$ and $t=1000$. We observe that the pattern is relatively flat in the directions of the RoI, and the maximum gain is obtained in the direction of the UE, i.e., $\theta_U = 70^\circ $.  Besides, it is also noted that to meet the orthogonality requirement of the sensing matrix, the beam pattern is disorderly in the directions of RoI at different time indexes. 

Then we investigate the receive beam pattern under different weight factor $\rho$. As shown in Fig. \ref{fig:fangxaingtu_rou}, the gain in the direction of the UE increases with $\rho$, while the gain in the direction of RoI is decreases accordingly. The results illustrate that the choice of $\rho$ is a trade-off between communication and imaging performance. 

In Fig. \ref{fig:fangxaingtu_bit}, we fix $\rho = 0.5$ and investigate the effect of $n_{bit}$ on the receive pattern. Compared with discrete phase model, a higher gain can be obtained in the direction of the UE. When $n_{bit}=1$, it is shown that there is a gain loss about 0.5 dB compared with the continuous phase. We also observe that the gain loss at the main lobe is only about 0.05dB when the phase shift is 2-bit quantized. 

\subsection{Performance of Uplink Joint Communication and Imaging}
In this subsection, we investigate the performance of the proposed message passing based joint communication and imaging method. We rearrange the scattering coefficients $\boldsymbol{\sigma}$ into a $8 \times 8$ matrix to make it intuitive. As shown in Fig. \ref{XDU} (a), (e) and (i), the letters X, D and U are respectively set to be the grand truth of $\boldsymbol{\sigma}$ in different scenarios of our simulations.

\begin{figure}
	\centering
	\includegraphics[width=0.9\linewidth]{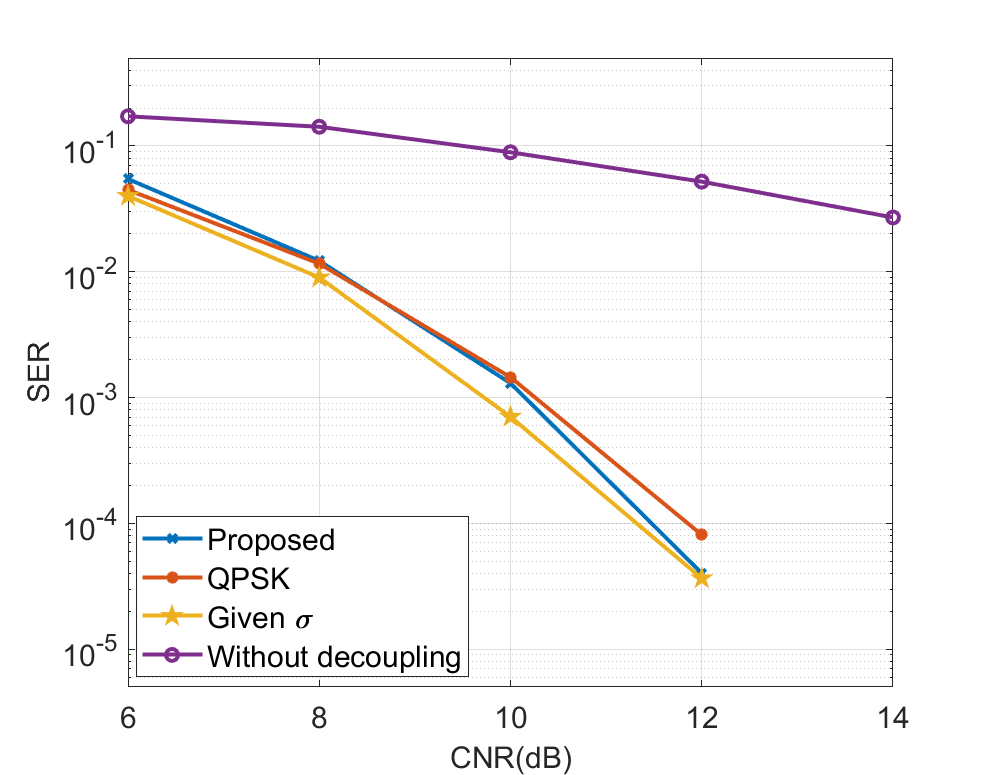}
	\caption{The SER performance of the proposed method versus CNR and INR when the CNR is 5 dB higher than the INR.}
	\label{fig:SER_X}
\end{figure}

\begin{figure}
	\centering
	\includegraphics[width=0.9\linewidth]{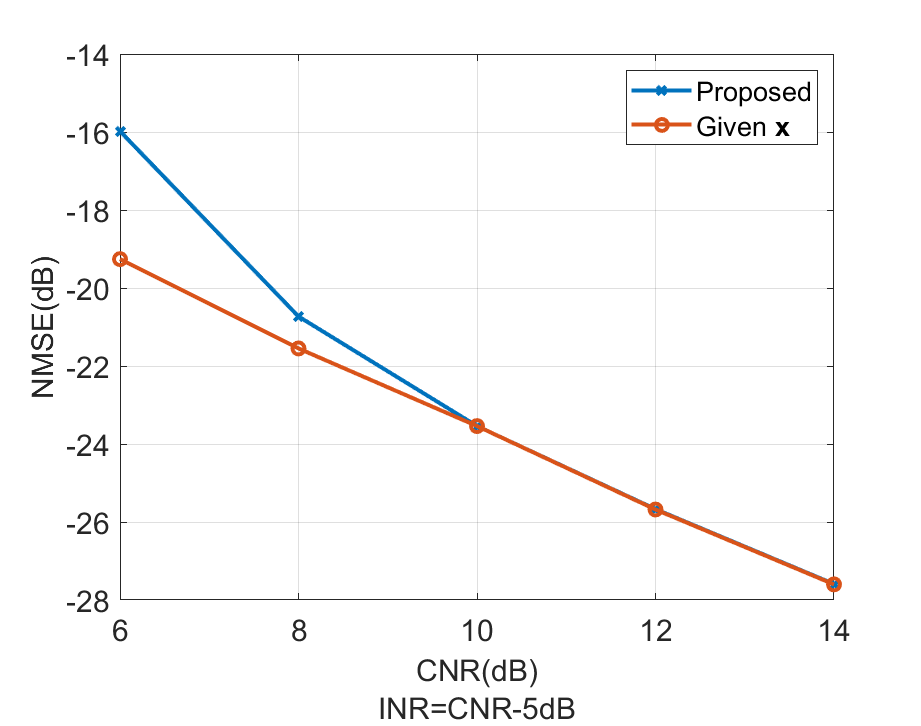}
	\caption{The NMSE performance of the proposed method versus CNR and INR when the CNR is 5 dB higher than the INR.}
	\label{fig:figx2}
\end{figure}

\begin{figure}
	\centering
	\includegraphics[width=0.9\linewidth]{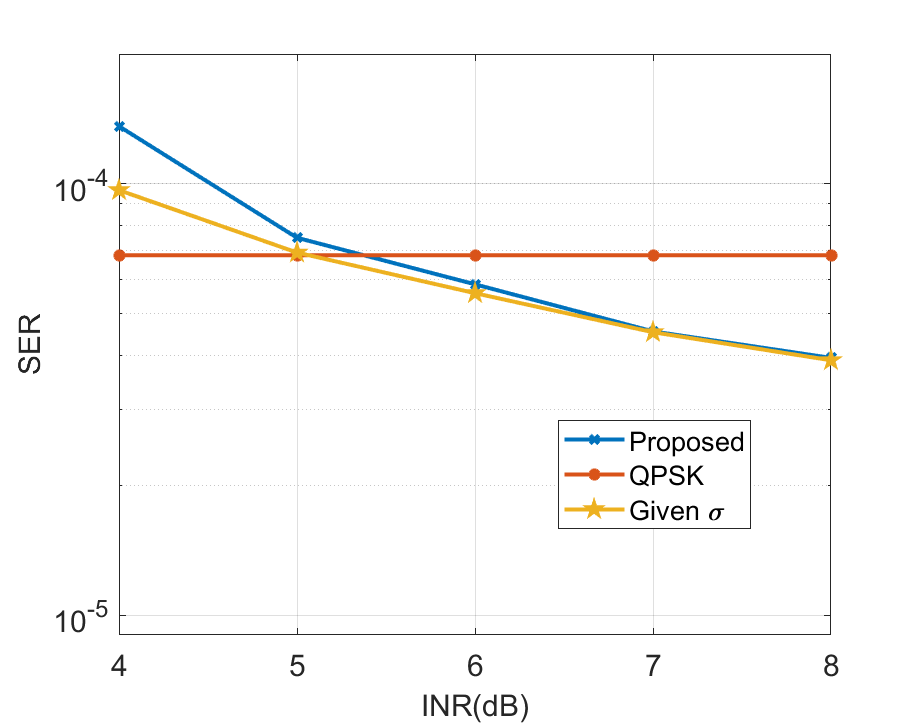}
	\caption{The SER performance of the proposed method versus INR when CNR = 12 dB.}
	\label{fig:figd1}
\end{figure}
\begin{figure}
	\centering
	\includegraphics[width=0.9\linewidth]{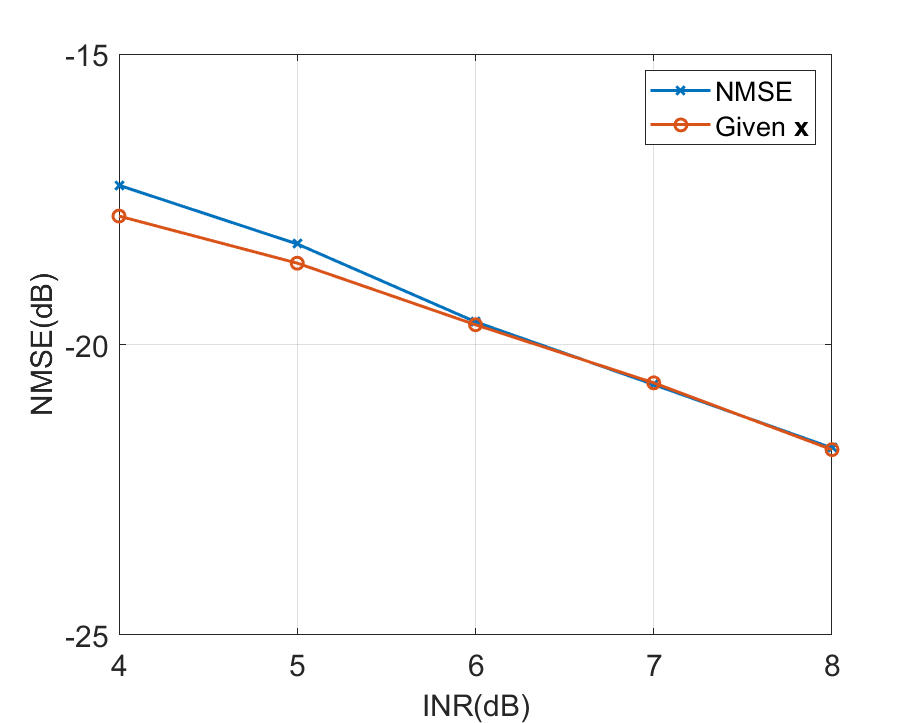}
	\caption{The NMSE performance of the proposed method versus INR when CNR = 12 dB.}
	\label{fig:figd2}
\end{figure}

\emph{ Scenario 1:} In the first scenario, we investigate communication and imaging performance versus the noise level. Note that the CNR and INR will change synchronous with the noise level, we fix CNR-INR = 5dB. The ground truth of $\boldsymbol{\sigma}$ is shown in Fig. \ref{XDU} (a). In Fig. \ref{fig:SER_X}, the SER performance versus CNR is plotted, along with three benchmarks to better illustrate the performance of the proposed method. It is not surprising that we observe the worst SER performance when we ignore the imaging echo directly, i.e., the echoes are not decouple. Besides, we pretend that there is no imaging echo, i.e., the received signals are simple uncoupled QPSK modulated signals, and we take the SER performance in this case as another benchmark. It is shown that under low CNR cases, the SER of the proposed method is higher than the QPSK signal. However, with the increase of CNR, the proposed method achieves a better SER performance than the QPSK signals. This result shows that under high SNR levels, the decoupled echoes are conducive to improving communication performance. Moreover, we also plot the SER performance under given $\boldsymbol{\sigma}$ as a low bound. It is shown that, with the increase of CNR, the SER of the proposed method can approaches this low bound gradually. 

To illustrate the imaging performance of the proposed method, the NMSE versus CNR is investigated in Fig.\ref{fig:figx2}, where the NMSE performance under given $\mathbf{x}$ is also included as a low bound. It can be seen that the proposed method can attain the low bound with the increase of CNR. To visually view the imaging results, we also show the $\boldsymbol{\sigma}_q$ in different NMSE levels in Fig. \ref{XDU} (b)-(d).

\emph{ Scenario 2:} In this scenario, we fix CNR = 12 dB and vary INR from 4 dB to 8 dB to evaluate the performance under different CNR levels. The ground truth of rearranged $\boldsymbol{\sigma}$ is given in Fig. \ref{XDU} (e). As shown in Fig. \ref{fig:figd1}, we observe that the communication performance improves with the increase of INR and gradually approaches the low bound in the case of given $\boldsymbol{\sigma}$. It also can be seen that, with the increase of INR, the SER performance  transcends the performance boundary of uncoupled QPSK signal, which also indicates that the imaging echo is utilized to enhance the communication performance. In addition, the NMSE performance is shown in Fig. \ref{fig:figd2}, along with the NMSE in the case of given $\mathbf{x}$ as a benchmark. In Fig. \ref{XDU} (f)-(h), we show the $\boldsymbol{\sigma}_q$ in different NMSE levels to view the imaging results visually. The ground truth of rearranged $\boldsymbol{\sigma}$ is given in Fig. \ref{XDU} (e). We can draw the conclusion that the imaging performance of the proposed method improves with the increase of CNR and it is able to reach the low bound asymptotically.

\begin{figure}
		\centering
		\includegraphics[width=0.9\linewidth]{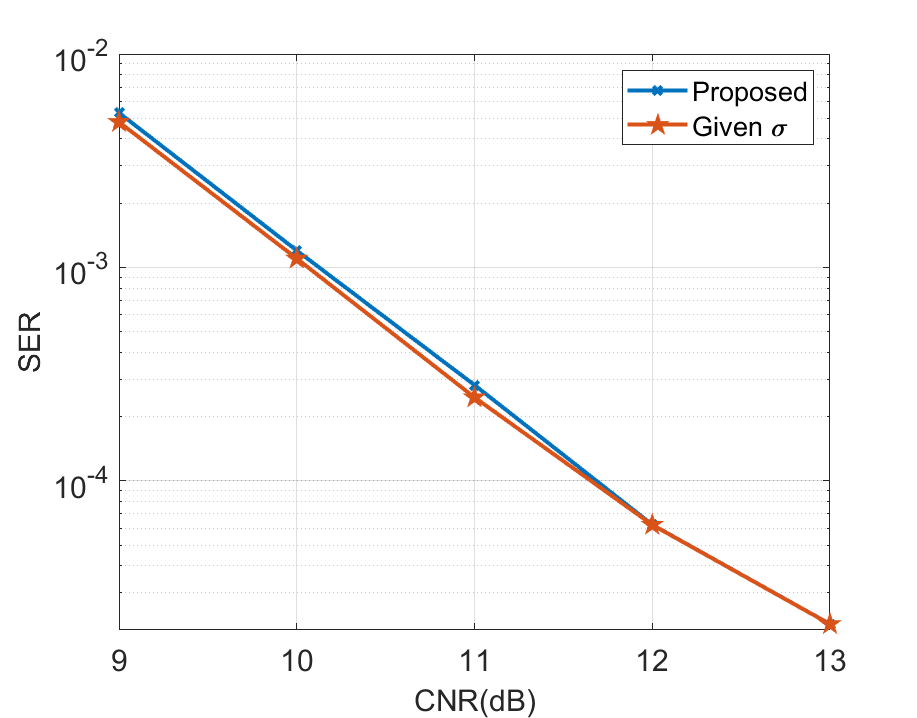}
		\caption{The SER performance of the proposed method versus CNR when INR = 6 dB.}
		\label{fig:figu1}
	\end{figure}

\begin{figure}
		\centering
		\includegraphics[width=0.9\linewidth]{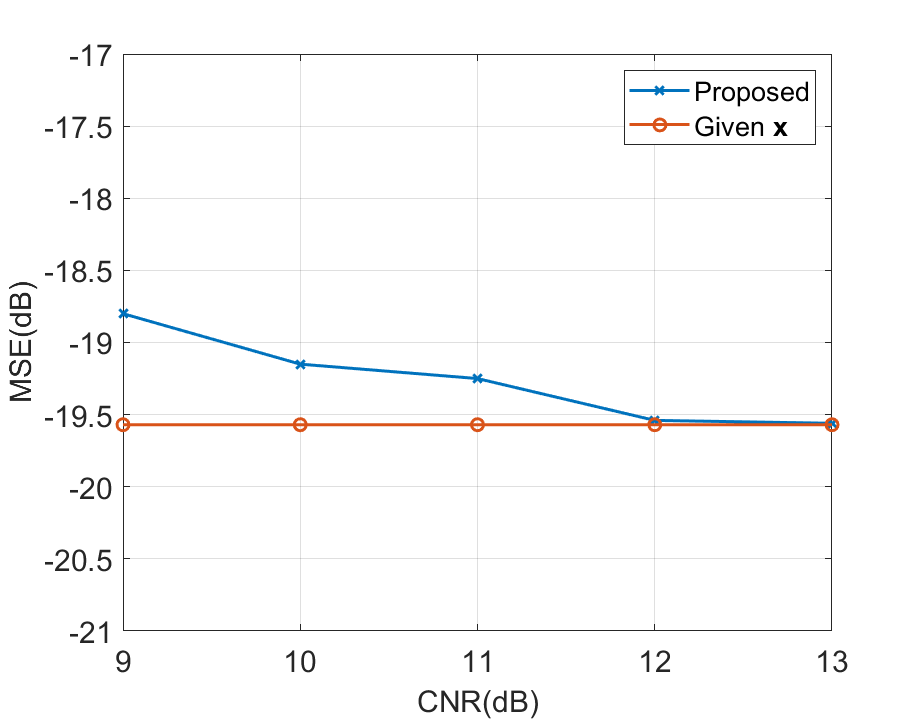}
		\caption{The NMSE performance of the proposed method versus CNR when INR = 6 dB.}
		\label{fig:figu2}
	\end{figure}

\emph{ Scenario 3:} Lastly, we fix INR = 6 dB and investigate the performance of the proposed method under different CNRs. The ground truth of the rearranged $\boldsymbol{\sigma}$ is given in Fig. \ref{XDU} (i), and the SER performance of the proposed method is shown in Fig. \ref{fig:figu1}. Besides, we consider the NMSE performance in the case of given $\boldsymbol{\sigma}$ as a benchmark. As depicted in Fig. \ref{fig:figu1}, the performance of the proposed method approaches the benchmark in a wide range of CNR. In Fig. \ref{fig:figu2}, we evaluate the imaging performance under different CNR levels. It can be noted that with the increase of CNR, the NMSE of the proposed method decreases and gradually approaches the NMSE performance under given $\mathbf{x}$.


		\section{Conclusion}
		We have presented a novel RIS-assisted uplink joint communication and imaging system in this article. A phase optimization problem is established with the consideration of both communication and imaging performance. Since the problem is non-convex and strongly coupled, we propose a back propagation based method, where temperature parameters and discrete prior are jointly utilized to solve it. After that, an efficient iterative message passing based method is proposed, cooperating the adaptive SBL to decouple echoes and achieve joint communication and imaging. Simulation results have demonstrated that the proposed phase design scheme can make a trade-off between communication and imaging functionality. It is also shown that the proposed method approaches the low bound asymptotically, and the communication performance can be enhanced by making full use of imaging echoes.


		\ifCLASSOPTIONcaptionsoff
		\newpage
		\fi

		
		
		
		%

		\bibliographystyle{IEEEtran}
		\bibliography{ref_TAP}

		


	\end{document}